\documentclass[12pt,journal,draftclsnofoot,onecolumn,twoside,letterpaper]{IEEEtran}

\usepackage{amsmath, amsthm, mathtools}
\usepackage{eepic, epic,color}
\usepackage{times}
\usepackage[latin1]{inputenc}
\usepackage{amsfonts}
\usepackage{psfrag}
\usepackage{amsbsy}
\usepackage{amssymb}
\usepackage[mathscr]{eucal}
\usepackage{latexsym}
\usepackage[T1]{fontenc}
\usepackage{comment}
\usepackage[dvips]{graphicx}
\usepackage{bm}
\usepackage{array}
\usepackage{verbatim,mathrsfs, cite}
\usepackage{algorithmic}
\usepackage{algorithm}
\usepackage{multirow}
\usepackage{xcolor}
\usepackage{caption}
\usepackage{subcaption}

\newcounter{lecnum}
\DeclarePairedDelimiter{\floor}{\lfloor}{\rfloor}

\newtheorem{defn}{Definition}
\newtheorem{theorem}{Theorem}
\newtheorem{lemma}{Lemma}
\newtheorem{proposition}{Proposition}

\theoremstyle{remark}
\newtheorem{rem}{Remark}  

\newcommand{\figwss}{0.41\columnwidth}
\newcommand{\figws}{0.51\columnwidth}

\newcommand{\figww}{0.74\columnwidth}
\newcommand{\figwww}{0.95\columnwidth}

\newcommand{\be}{\begin{equation}}
\newcommand{\ee}{\end{equation}}
\newcommand{\ba}{\begin{align}}

\newcommand{\abs}[1]{\lvert#1\rvert}

\def\B(#1){\hbox{\mathbf$#1$}}
\def\C(#1){{\cal #1}}

\graphicspath{{./Figure/}}

\begin{document}

\title{\huge A Dynamic Network Formation Model for Understanding Bacterial Self-Organization into Micro-Colonies}

\author{
Luca~Canzian$^\diamond$,
Kun~Zhao$^\S$, 
Gerard~C.~L.~Wong$^{\S,\dag,\ddag}$,
and~Mihaela~van~der~Schaar$^\diamond$
{ \small
\\
$^\diamond$Department of Electrical Engineering, UCLA, Los Angeles CA 90095, USA.
\\
$^\S$Department of Bioengineering, UCLA, Los Angeles CA 90095, USA.
\\
$^\dag$Department of Chemistry and Biochemistry, UCLA, Los Angeles CA 90095, USA. 
\\
$^\ddag$California NanoSystems Institute, UCLA, Los Angeles CA 90095, USA. 
\\
}
}

\maketitle

\begin{abstract}
We propose a general parametrizable model to capture the dynamic interaction among bacteria in the formation of micro-colonies.
micro-colonies represent the first social step towards the formation of structured multicellular communities known as bacterial biofilms, which protect the bacteria against antimicrobials.
In our model, bacteria can form links in the form of intercellular adhesins (such as polysaccharides) to collaborate in the production of resources that are fundamental to protect them against antimicrobials. 
Since maintaining a link can be costly, we assume that each bacterium forms and maintains a link only if the benefit received from the link is larger than the cost, and we formalize the interaction among bacteria as a dynamic network formation game.
We rigorously characterize some of the key properties of the network evolution depending on the parameters of the system.
In particular, we derive the parameters under which it is guaranteed that all bacteria will join micro-colonies and the parameters under which it is guaranteed that some bacteria will not join micro-colonies.
Importantly, our study does not only characterize the properties of networks emerging in equilibrium, but it also provides important insights on how the network dynamically evolves and on how the formation history impacts the emerging networks in equilibrium. 
This analysis can be used to develop methods to influence on-the-fly the evolution of the network, and such methods can be useful to treat or prevent biofilm-related diseases. 
\end{abstract}


\section{Introduction} 
\label{sec:intro} 

Bacteria have a tendency to attach to surfaces and self-organize into \emph{micro-colonies}, which represent the first step toward the formation of \emph{biofilms}.
Biofilms are surface associated communities that are encased within an extracellular matrix, which can function as a structural scaffold and as a protective barrier to antimicrobials  \cite{Stoodley2002, Parsek03}.
In fact, biofilm communities exhibit enhanced antibiotic tolerance and biofilm infections are notoriously difficult to treat \cite{Parsek03, Costerton99, Drenkard03}. 

Key components of the biofilm extracellular matrix are the \emph{exopolysaccharides} 
, which are responsible for a wide range of functions involving cell-to-surface and cell-to-cell interactions \cite{Stoodley2002}, and can impart resistance to antibiotics \cite{Gerard11}.
Our prior study \cite{Gerard13}, however, has shown that some specific types of exopolysaccharides play also an active roles in the early stage organization of micro-colonies and biofilms. 
A phenomenological model of the exopolysaccharides impact on the dynamics of micro-colonies and biofilm development, one using simple assumptions and well-controlled approximations, would provide crucial guidance to our understanding of biofilms and the design of biofilm therapeutic strategies. 

In this work, we focus on the implications of having the simplest type of adhesion molecule between bacteria, a cell-to-cell adhesin that forms a linkage between two cells. 
We propose a general parametrizable model, built based on experimental evidence obtained from various studies including our own prior works \cite{Gerard11, Gerard13}, to capture the dynamic interaction among bacteria in the formation of micro-colonies.
In our model, bacteria move along a surface and produce \emph{resources} (a generalized model for inter-cell adhesions, like polysaccharides), which spread in space to an extent that we can control in the model, and give a benefit to all bacteria which get access to them. 
When two bacteria approach each other, in order to benefit from the resources produced by the other, each of them can decide to stop moving. 
We call this process ``link'' formation and we state that a link is maintained between two bacteria if they remain close. 
Bacteria can also link with bacteria that are already linked to other bacteria, as well as can break existing links.

To develop a general model, we abstract in this paper from the mobility and motility models of bacteria (which depend on the particular strain of bacteria), and from the geometric properties of the surface and the positions of bacteria, and we assume that the meetings among bacteria are governed by a random process, like in \cite{Watts01}.  
We consider a discrete time model and assume that in each time slot each bacterium can break some of its links, and a non-linked (singleton) bacterium is matched with a certain probability with another (linked or non-linked) bacterium and they decide whether to form a link. 
Since in our prior study we observed that areas rich of resources act as a \emph{signaling mechanism} that attracts bacteria toward them \cite{Gerard13}, we assume that the probability of being matched with a bacterium that has high connectivity (many links), and hence it is located in an area rich of resources, is higher than the probability of being matched with a bacterium that has low connectivity.

We consider a population including bacteria in two different physiological states, which we refer to as \emph{types}.
\emph{High type} bacteria increase their resource production rate when linked to other bacteria, for example \cite{Gerard13} shows that the resource production rates of some bacteria increase when they are located in areas rich of resources.
\emph{Low type} bacteria are considered to have constant resource production rate regardless of their links \cite{Gerard13}.
Moreover, we make the assumption that bacteria are \emph{selfish and myopic} \cite{Velicer03}, i.e., they choose whether to form or break a link such that
they maximize their immediate utilities. 

We will study two limiting cases. 
In the case of \emph{complete information} bacteria know in advance the immediate utilities they will obtain by forming links with other specific bacteria. 
In the case of \emph{incomplete information} bacteria know the utilities associated to a link only after they have formed it. 
In the context of microbiology, if bacteria have a long time to sample the environment and integrate inputs (which corresponds to slow motility compared to gene expression and protein synthesis response times), or if the environmental conditions are slowly-varying in time, then bacteria are well-described by the first limit. 
In contrast, if  bacteria move quickly relative to their response times or if the environmental conditions vary significantly in time, then the latter limit would be more appropriate. 
For example, something as simple as nutritional conditions can impact motility and biochemical signaling \cite{Shrout06}.


As a main contribution of this work, we characterize the possible evolutions of the micro-colonies formed by bacteria depending on the system parameters. 
In particular, we define a \emph{stable network} as a network in which all bacteria belong to some micro-colonies, and we analytically derive the conditions on the system parameters, in both complete and incomplete information settings, under which a stable network emerges, under which a stable network does not emerge, and under which the emergence of a stable network is determined by chance.

\begin{figure}
     \centering
          \includegraphics[width=\figww]{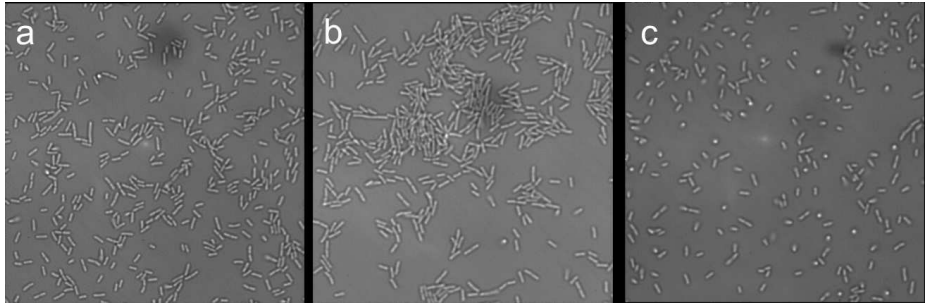}
\caption{Variation of Pseudomonas aeruginosa communities under different conditions: a) small micro-colonies formed by wild types, b) large micro-colonies formed by mutants that overproduce polysaccharide Psl, c) no micro-colonies formed by mutants that cannot produce Psl.}
\label{fig:para_intro}
\end{figure}

The types of behavior predicted by our model are observed in real-world experiments. 
For example, Fig. \ref{fig:para_intro} shows the variation of the communities formed by Pseudomonas aeruginosa -- a bacterium widely used in biofilm research \cite{Drenkard03, Gerard11, Gerard13} -- under different conditions. 
Each of the three images is a snapshot of the bacterial movement on a $67 \times 67 \mu m$ glass surface.
Each snapshot is taken with an Olympus microscope when the total number of bacterial visits on the surface reaches $1.5 \times 10^6$.  
For this experiment we use the wild type Pseudomonas aeruginosa strain $PAO1$ (Fig. \ref{fig:para_intro}.a), and its isogenic mutants $\Delta Ppsl/PBAD-psl$ (Fig. \ref{fig:para_intro}.b) and $\Delta pslD$ (Fig. \ref{fig:para_intro}.c).
The mutant $\Delta Ppsl/PBAD-psl$ with $1\%$ arabinose added into the medium produces larger amounts of polysaccharide Psl than the wild type strain $PAO1$, whereas the mutant $\Delta pslD$ cannot produce Psl.
Fig. \ref{fig:para_intro} shows that large micro-colonies are formed by mutants that overproduce Psl; instead, no micro-colonies are formed by mutants that cannot produce Psl.

The results presented in this manuscript represent an important step toward the derivation of bacterial behavior models, which can help us answer several enabling questions such as: How do combinations of motility, polysaccharide production, and antibiotics influence bacterial decisions to form micro-colonies and eventually biofilms that are inherently more tolerant to antibiotics?
Present strategies to treat biofilm-related diseases are based solely on killing bacteria, which result in a large selection pressure to evolve drug-resistant strains. 
Conversely, a dynamic model of network formation and micro-colony development allows to foresee how a bacterial community responds to a complex set of stimuli, and whether bacterial community decisions can be influenced. 
Since our model is dynamic and incorporates tunable parameters, it allows for specific microbiological experiments to be designed. 
Moreover, our general model can be calibrated for various strains of bacteria and types of linkage-generating adhesions, such as polysaccharides. 
Finding the inputs that will disassemble a micro-colony in a controlled manner will have a transformative impact on biofilm therapeutics.  

The rest of the paper is organized as follows. 
Section \ref{sec:rel} reviews the existing literature in nano-scale communications and network formation. 
Section \ref{sec:mod} describes our model and introduces the basic concepts necessary to study how micro-colonies form.  
Section \ref{sec:dyn} formalizes the interaction of the bacteria as a dynamic network formation game.
Section \ref{sec:ana} analyzes the dynamic network formation game. 
Section \ref{sec:sim} presents several illustrative results aimed to understand the essential characteristics of micro-colonies formation and their dependence on the key parameters of the system. 
Section \ref{sec:con} concludes the paper.


\section{Related Works}
\label{sec:rel}


\subsection{Molecular communication} 

Molecular communication \cite{Moore09}
is a nanoscale communication paradigm that has emerged recently, that enables engineered biological nanomachines to exchange information with the natural biological nanomachines which form a biological system. 
Distinct from the current telecommunication paradigm, molecular communication uses molecules as the carriers of information.
Many works propose and study models for molecular communication \cite{Moore09, Pierobon11, Pierobon13, Eckford11, Eckford12_1, Atakan12}.
For example, \cite{Pierobon11} proposes a model for the reception noise, \cite{Moore09} studies some approaches to reduce the noise, \cite{Pierobon13, Eckford11, Eckford12_1} investigate the channel capacity and achievable rates, and \cite{Atakan12} proposes a synchronization-free molecular communication scheme.

Our work is clearly different from this literature because it does not focus on methods to enable the exchange of information among static nanomachines.
Our research focuses on deriving network formation game models to analyze the dynamic formation of micro-colonies of bacteria. 


\subsection{Network Science} 

There is a renowned literature that first studies network formation as the result of strategic interaction among a group of self-interested agents \cite{Jackson96, Goyal00, Johnson00, Watts01}. 
In these papers, it is common to assume \emph{agent homogeneity}. 
This is the strongest form of \emph{complete information}, in the sense that agents do not only know their exact utilities from linking to others, but are also aware that the utilities are solely determined by the network topology, and that the agents' identities have no role in affecting utility characteristics.  
We differ from these works because in our paper bacteria are \emph{heterogeneous} in utilities they provide to and receive from others, and they are \emph{myopic}, i.e., they only care about their immediate utility when making decisions.  

In the network formation literature that considers agent heterogeneity, the concept ``heterogeneity'' is interpreted from various angles: as differentiated failure probabilities for different links \cite{Haller03}, as differences in values obtained from links and costs for forming links across agents \cite{Galeotti06, Goyal06}, as different amounts of valuable information produced endogenously \cite{Goyal10, Zhang12}, as different endogenous effort levels \cite{Koenig12}, or as agent-specific resource holding amounts \cite{Zhang13}. 
Despite such heterogeneity, these works assume \emph{complete information}, i.e., that agents know their exact utilities from linking to others. 
We differ from these works because in our paper bacteria are myopic and we analyze also the \emph{incomplete information} setting, in which a bacterium does not know the type of a bacterium it meets and whether this bacterium is connected to other bacteria. 


Importantly, most of the above cited works study the properties of a network after it has achieved a stable point, but they do not analyze \emph{if} and, possibly, \emph{how} the network becomes stable.
A key merit of our work, as opposed to the above literature, is that we study the dynamic evolution of the network, which has two fundamental advantages. 
First, we are able to predict under which conditions a stable network does, does not, or may emerge.
Second, the understanding of the network evolution process allows to develop methods to \emph{influence on-the-fly the evolution} of the network, and such methods can be useful to treat or prevent biofilm-related diseases.

\section{System Model} 
\label{sec:mod}

We consider a population of bacteria that are distributed on a surface. 
We denote by $K$ the number of bacteria in the surface and by $\mathcal{K} = \{ 1, \ldots,  K \}$ the set of bacteria. 
In our model, bacteria move along the surface, produce resources, and form new links or break existing links with other bacteria.
The \emph{resources} produced by the bacteria are secreted polysaccharides. 
At present, our understanding of the chemical nature of these polysaccharides, and therefore the wetting properties of such polysaccharides on surfaces, is incomplete. 
To have the most general possible model, we assume that part of the resource is adhered to the cell body \cite{Gerard11}, and part of it \emph{wets} the surface via spreading. 
In Subsection \ref{sec:ut} we will introduce a tunable parameter, the \emph{spread factor} $\delta$, to fit the properties of specific polysaccharides once their physical properties are known. 

We say that there is a \emph{link} between two bacteria $i$ and $j$ if their distance is below a certain threshold and they intentionally decide not to move and maintain such a distance.
The images in \cite{Gerard11} suggest that such distance threshold is about $0.3$ microns, i.e., $\frac{1}{10}$ the size of a typical cell.\footnote{Notice that we abstract from geometric concepts. Our model can take into account the effects of a lower (higher) distance by increasing (decreasing) the spread factor and decreasing (increasing) the probability that two bacteria meet.}
A link is formed and maintained \emph{bilaterally}, i.e., \emph{both} bacteria must agree to form and maintain the links.
Indeed, if a bacterium does not want to maintain a link it can leave that specific location and break the link.  
Moreover, a link is \emph{undirected}, because a link between $i$ and $j$ automatically implies a link between $j$ and $i$. 
If two bacteria are linked, because of their proximity, each of them can exploit part of the resources produced but not used by the other bacterium, i.e., two bacteria have a mutual benefit to stay close to each other. 

We consider a population including bacteria in two different physiological states, which we refer to as \emph{types}.
Different types of bacteria differ in the \emph{resource production rate} they adopt when they are linked with other bacteria: \emph{high type} bacteria ($H$) increase their resource production rate when linked to other bacteria, whereas \emph{low type} bacteria ($L$) are considered to have constant resource production rate regardless of their links \cite{Gerard13}.
We denote by $K(L)$ and $K(H) = K-K(L)$ the number of low and high type bacteria, respectively, and by $\rho_L \triangleq \frac{K(L)}{K}$ and $\rho_H \triangleq \frac{K(H)}{K}$ the ratio of low and high type bacteria, respectively. 
Also, we denote by $t_i \in \{L, H\}$ 
the type of bacterium $i$, $i \in \mathcal{K}$, and by $\mathbf{t} = (t_1, \ldots, t_K)$ the type profile. 

In this paper, we divide the time into equal slots and we characterize the history of interactions among bacteria and its influence on the formation of micro-colonies.
We write $g_{ij}^{(n)} = 1$ if bacteria $i$ and $j$ are linked at the beginning of time slot $n$, and $g_{ij}^{(n)} = 0$ otherwise.
We formally define the \emph{network} $\mathcal{G}^{(n)}$ as the set of all bacteria that are linked at the beginning of the $n$-th time slot, $\mathcal{G}^{(n)} \triangleq \{ (i,j): g_{ij}^{(n)} = 1 \}$, and the pair $\left( \mathcal{K}, \mathcal{G} \right)$ represents a \emph{graph} \cite{WestGraphBook}.
The network at the end of time instant $n$ is denoted by $\mathcal{G}^{(n+1)}$, because it corresponds to the network at the beginning of the next time instant.
It is also useful to define the \emph{intermediate network} $\overline{\mathcal{G}}^{(n)}$ in time instant $n$, as described in Section \ref{sec:dyn} this represents an intermediate step between $\mathcal{G}^{(n)}$ and $\mathcal{G}^{(n+1)}$. 

We define the length $\ell_{ij}^{(n)} $ of a link $(i, j)$ at time slot $n$ as the number of slots since the link has been formed, i.e., $\ell_{ij}^{(n)} \triangleq n - \max \{ m \leq n : g_{ij}^{(m)} = 0 \}$.
Since there is a reaction time between when bacteria detect environmental cues and when they response to them \cite{Guzman95}, we define the \emph{minimum link length} $\ell_{min} \geq 1$, meaning that a link $(i,j)$ cannot be broken if $\ell_{ij}^{(n)} < \ell_{min}$.

Given a network $\mathcal{G}$ (that may be either $\mathcal{G}^{(n)}$ or $\overline{\mathcal{G}}^{(n)}$), we define the \emph{set of $i$'s neighbors} as the set of bacteria to which bacterium $i$ is linked to, $\mathcal{N}_i \triangleq \{ j \in \mathcal{K} : g_{ij} = 1 \}$, and we say that bacteria $i$ and $j$ are connected, denoted by $i \xleftrightarrow{\mathcal{G}} j$, if $(i,j) \in  \mathcal{G}$ or there are $k$ bacteria such that $(i,i_1), (i_1,i_2), \ldots, (i_{k},j) \in \mathcal{G}$. 
We define the \emph{distance} $d_{ij}$ between two connected bacteria $i$ and $j$ as the smallest number of links between $i$ and $j$.
We say that bacterium $i$ is \emph{singleton} if $(i,j) \notin \mathcal{G}$, $\forall \, j \neq i$, and we denote by $\mathcal{K}_S$ 
the set of singleton bacteria. 

$\left( \underline{\mathcal{K}}, \underline{\mathcal{G}} \right)$ is a subgraph of $\left( \mathcal{K}, \mathcal{G} \right)$ 
if $\underline{\mathcal{K}} \subseteq \mathcal{K}$ and $\underline{\mathcal{G}}$ contains all the original links among the bacteria in $\underline{\mathcal{K}}$. 
A \emph{component} $\mathcal{C}$ 
of a graph is a subgraph in which any two bacteria are connected to each other and which is connected to no additional bacteria in the original graph. 
We abuse notation and write $i \in \mathcal{C} $ if bacterium $i$ belongs to the set of bacteria defined by the component $\mathcal{C} $.
The \emph{size} $\abs{\mathcal{C}}$ 
of a component $\mathcal{C}$ is the number of bacteria belonging to the component, whereas the diameter $D_{\mathcal{C}} \triangleq \max_{i,j \in \mathcal{C}} d_{ij}$ is the maximum distance between two bacteria $i$ and $j$ belonging to $\mathcal{C}$.
Note that a singleton bacterium is itself a component with size $1$ and diameter $0$, and each bacterium $i \in \mathcal{K}$ belongs to one and only one component.

\vspace{-0.1cm}
\section{Dynamic Network Formation Game} 
\label{sec:dyn}

In this section we formalize the interaction of the bacteria as a \emph{dynamic network formation game} \cite{Watts01}, 
in which bacteria are assumed to be \emph{myopic}, i.e., they select their actions to maximize their immediate utilities. 
Moreover, we define the concepts of \emph{micro-colony} and {stable network}, which will be used in Section \ref{sec:ana} to analyze how the network $\mathcal{G}^{(n)}$ evolves in time.

\vspace{-0.2cm}
\subsection{Utility structure}
\label{sec:ut} 

Antibiotic tolerance develops very early in the formation of a biofilm, on the order of $1$-$3$ hours after a community initiates \cite{Mah03}. 
Our prior study has attributed this to polysaccharides \cite{Gerard11}.  
Hence, we assume that a bacterium $i$ obtains a benefit, quantified by a utility  function, whenever it links to another bacterium $j$, because it exploits part of the resources produced but not used by $j$.
Given the network $\mathcal{G}^{(n)}$ in time instant $n$, we define bacterium $i$'s utility as follows:
\ba
&u_i (\mathbf{t}, \mathcal{G}^{(n)}) \triangleq \left\{\begin{array}{ll}
0 & \mbox{if} \; i \in \mathcal{K}_S^{(n)} \\
\sum_{j \leftrightarrow i} \delta^{d_{ij}-1} f(t_j) - c(t_i)  & \mbox{otherwise}
\end{array} \right.
\label{eq:ut}
\end{align}
$f(t_j)>0$ represents the benefit that bacterium $i$ receives from bacterium $j$ (having type $t_j$) it is linked with, 
$\delta \in (0,1)$ is the \emph{spread factor}, such that bacterium $i$ can also benefit from a bacterium $j$ it is not directly linked with, but such a benefit decreases exponentially in their distance, and
$c(t_i) \geq 0$ is a cost to pay to be part of a non-singleton component.
Since high type bacteria increase their resource production rate when linked to other bacteria and low type bacteria always adopt a constant production rate, we consider $f(H) > f(L)$ and $c(H) > c(L) = 0$. 

The benefit that a bacterium $i$ achieves when forming a link with a bacterium $j$ is more attractive if $j$ is already connected to many bacteria. 
We refer to this effect as \emph{increasing returns to link formation}.
This means that it is desirable and efficient for bacteria to be part of components having large sizes. 
This is coherent with the experimental observation that biofilm exhibit enhanced antibiotic tolerance \cite{Drenkard03}. 
However, to understand under which conditions \emph{individual bacteria} have an incentive to begin such a formation process, we need to formalize the interaction among bacteria in each time slot as a game (Subsection \ref{sec:game}), define the equilibrium concepts of this interaction (Subsection \ref{sec:eq}), define stable states for the network (Subsection \ref{sec:micro}), and study how the network can evolve (Section \ref{sec:ana}). 

\vspace{-0.3cm}
\subsection{The game} 
\label{sec:game} 

The interaction among bacteria is modeled as follows.
At time slot $1$, bacteria form an empty network, i.e., $\mathcal{G}^{(1)} = \emptyset$.
In a generic time slot $n$, the following events happen sequentially:
\begin{itemize}
\item[\textbf{1}.] for each link $(i,j) \in \mathcal{G}^{(n)}$ such that $\ell_{ij}^{(n)} \geq \ell_{min}$, bacteria $i$ and $j$ select whether to break the link $(i, j)$. Denote by $a_{ij}^{(n)} \in \{0,1\}$ and $a_{ji}^{(n)} \in \{0,1\}$ the choice of bacteria $i$ and $j$, respectively, where $1$ ($0$) means that the bacterium wants to maintain (break) the link. Since links are maintained bilaterally, the new network after this interaction is $\overline{\mathcal{G}}^{(n)} \triangleq \{ (i,j): \overline{g}_{ij}^{(n)} = 1 \}$, where $\overline{g}_{ij}^{(n)} \triangleq \min \{ a_{ij}^{(n)} , a_{ji}^{(n)} \}$;
\item[\textbf{2}.] with a certain probability one singleton bacteria $i \in \overline{\mathcal{K}}_S^{(n)}$
approaches another bacteria $j$ (singleton or non singleton), this event is denoted by $(i,j) \in \mathcal{E}^{(n)}$, and they decide whether to form a link. 
Denote by $s_{ij}^{(n)} \in \{0,1\}$ and $s_{ji}^{(n)} \in \{0,1\}$ the choice of bacteria $i$ and $j$, respectively, where $1$ ($0$) means that the bacterium wants to form (not to form) the link. Since the link is formed bilaterally, the new network after this interaction is $\mathcal{G}^{(n+1)} \triangleq \overline{\mathcal{G}}^{(n)}$ if $g_{ij}^{(n+1)} \triangleq \min \{ s_{ij}^{(n)} , s_{ji}^{(n)} \} = 0$, and $\mathcal{G}^{(n+1)} \triangleq \overline{\mathcal{G}}^{(n)} \bigcup \{ (i , j) \}$ if $g_{ij}^{(n+1)} = 1$;
\item[\textbf{3}.] each bacterium $i$ receives the utility $u_i (\mathbf{t}, \mathcal{G}^{(n+1)})$.
\end{itemize}

The meeting among bacteria in time step \textbf{2} is modeled as follows.
With probability $\gamma(\abs{\overline{\mathcal{K}}_S^{(n)}})$, increasing in the number of singleton bacteria in $\overline{\mathcal{G}}^{(n)}$, one singleton bacterium $i$ is picked uniformly in $\overline{\mathcal{K}}_S^{(n)}$ and is matched with one bacterium $j$, $ j \neq i $. 
Notice that we do not allow the formation of multiple links during the same time slot. An interpretation for this is that the considered time slot is so short that the probability that more than one pairs of bacteria meet is negligible compared to the probability that only one pair of bacteria meets.

Since in our prior study we observed that areas rich of resources act as a \emph{signaling mechanism} that attracts bacteria toward them \cite{Gerard13}, we assume that the probability of being matched with a bacterium $j$ that has high connectivity (many links), and hence it is located in an area rich of resources, is higher than the probability of being matched with a bacterium that has low connectivity.
Specifically, bacterium $j$ is drawn from the distribution
\ba
p_j = \dfrac{ h \left( \abs{\mathcal{N}_j^{(n)}} \right) }{\sum_{k \neq i }  h \left( \abs{\mathcal{N}_k^{(n)}} \right) } \;\;, \;\; j \in \mathcal{K} \;\;, \;\;  j \neq i \;\; , \nonumber
\end{align} 
where the \emph{signaling mechanism} $h : \mathbb{N} \to \Re^+$ is a positive non-decreasing function. 

\subsection{Equilibrium concepts for complete and incomplete information settings} 
\label{sec:eq} 

We consider the interaction among bacteria in two different scenarios.
In the \emph{complete information setting} bacteria know in advance the immediate utilities they will obtain by forming links with other specific bacteria.
This implies that, when bacterium $i$ approaches bacterium $j$, if $j$ is singleton then $i$ is able to recognize its type (\cite{Gibbs08} shows that some bacteria have this ability), whereas if $j$ is not singleton then $i$ is able to estimate the amount of resource produced by the component $j$ belongs to.
We define the following equilibrium concept for the complete information game.

\begin{defn}
An action profile $\mathbf{a}^{(n)} = \left( \{ a_{ij} \}_{(i,j) \in \mathcal{G}^{(n)}: \ell_{ij}^{(n)} \geq \ell_{min}} , \{ s_{\tilde{i}\tilde{j}} \}_{(\tilde{i},\tilde{j}) \in \mathcal{E}^{(n)}} \right)$ in time instant $n$ is a myopic equilibrium in the complete information setting if and only if, $\forall \, i \in \mathcal{K} : \mathcal{N}_i^{(n)} \neq \emptyset$, $\forall \, j \in \mathcal{N}_i^{(n)} : \ell_{ij}^{(n)} \geq \ell_{min}$, $\forall \, \hat{a}_{ij} \in \{0, 1\}$, and $\forall \, (\tilde{i},\tilde{j}) \in \mathcal{E}^{(n)}$, the following conditions are satisfied
\begin{itemize}
\item[C1] $u_i \left( \mathbf{t}, \mathcal{G}^{(n)} - \{ (i,j) : a_{ij}^{(n)} =0 \} \right) \geq 
u_i \left( \mathbf{t}, \mathcal{G}^{(n)} - \{ (i,j) : \hat{a}_{ij}^{(n)} =0 \} \right)  ,$
\item[C2] if $\exists \, j : a_{ij}^{(n)} =0$ then $u_i \left( \mathbf{t}, \mathcal{G}^{(n)} - \{ (i,j) : a_{ij}^{(n)} =0 \} \right) > 
u_i \left( \mathbf{t}, \mathcal{G}^{(n)} \right)$ , 
\item[C3] $s_{\tilde{i} \tilde{j}}^{(n)} =1$ if and only if $u_{\tilde{i}} \left( \mathbf{t}, \overline{\mathcal{G}}^{(n)} \cup (\tilde{i},\tilde{j}) \right)  \geq  u_{\tilde{i}} (\mathbf{t}, \overline{\mathcal{G}}^{(n)} ) $.
\end{itemize}
\end{defn}

Condition C1 states that bacterium $i$ selects the actions $a_{ij}^{(n)}$, $j \in \mathcal{N}_i^{(n)} $, i.e., which links to maintain, to maximize (a posteriori) the utility $u_i (\mathbf{t}, \mathcal{G}^{(n)})$ received at the end of the last time slot. 
Condition C2 states that bacterium $i$ prefers to maintain the links instead of breaking them if the resulting utility is the same.
Condition C3 states that bacterium $\tilde{i}$ that approaches bacterium $\tilde{j}$ selects the action $s_{\tilde{i}\tilde{j}}^{(n)}$, i.e., whether to form a new link with $\tilde{j}$, to maximize the utility $u_{\tilde{i}}  (\mathbf{t}, \mathcal{G}^{(n+1)})$ it will obtain at the end of the current slot. 

In the \emph{incomplete information setting} bacteria do not know the utilities they will obtain by forming a link, because they are not able to detect the type of the other bacteria and the amount of resource generated by the component the other bacteria belong to.
In this case, we assume that bacteria \emph{always} form new links, i.e., $s_{\tilde{i}\tilde{j}}^{(n)} = 1$, $\forall \, (\tilde{i},\tilde{j}) \in \mathcal{E}^{(n)}$, and that each bacterium $i$ selects the actions $a_{ij}^{(n)}$, $j \in \mathcal{N}_i^{(n)} $, i.e., which links to maintain, to maximize (a posteriori) the utility $u_i (\mathbf{t}, \mathcal{G}^{(n)})$ received at the end of the last time slot.
Hence, the equilibrium concept for the incomplete information game is defined as follows.

\begin{defn}
An action profile $\mathbf{a}^{(n)} = \left( \{ a_{ij} \}_{(i,j) \in \mathcal{G}^{(n)}: \ell_{ij}^{(n)} \geq \ell_{min}} \right)$ in time instant $n$ is a myopic equilibrium in the incomplete information setting if and only if, $\forall \, i \in \mathcal{K} : \mathcal{N}_i^{(n)} \neq \emptyset$, $\forall \, j \in \mathcal{N}_i^{(n)} : \ell_{ij}^{(n)} \geq \ell_{min}$, $\forall \, \hat{a}_{ij} \in \{0, 1\}$, and $\forall \, (\tilde{i},\tilde{j}) \in \mathcal{E}^{(n)}$, the following conditions are satisfied
\begin{itemize}
\item[C4] $u_i \left( \mathbf{t}, \mathcal{G}^{(n)} - \{ (i,j) : a_{ij}^{(n)} =0 \} \right) \geq 
u_i \left( \mathbf{t}, \mathcal{G}^{(n)} - \{ (i,j) : \hat{a}_{ij}^{(n)} =0 \} \right)  ,$
\item[C5] if $\exists \, j : a_{ij}^{(n)} =0$ then $u_i \left( \mathbf{t}, \mathcal{G}^{(n)} - \{ (i,j) : a_{ij}^{(n)} =0 \} \right) > 
u_i \left( \mathbf{t}, \mathcal{G}^{(n)} \right)$.
\end{itemize}
\end{defn}

\begin{proposition}
In both complete and incomplete information settings, in each time instant $n$ the myopic equilibrium exists and is unique. 
\label{prop:uniq}
\end{proposition}

\begin{IEEEproof}
A myopic equilibrium for the incomplete information setting can be obtained maximizing, for each bacterium $i$, the left side of condition C4 with respect to the actions $a_{ij}^{(n)}$, $j \in \mathcal{N}_i^{(n)}$. 
Since the action space if finite, a solution exists. 
Moreover, each solution must satisfy either $a_{ij}^{(n)}=0$, $\forall \, j \in \mathcal{N}_i^{(n)}$, or $a_{ij}^{(n)}=1$, $\forall \, j \in \mathcal{N}_i^{(n)}$.
Indeed, maintaining all links is strictly better than maintaining only a subset of them, because the cost to pay for a single link is equal to the cost to pay for multiple links.
Hence, there are only two possible solutions to maximize the left side of condition C4.
If they are equivalent in terms of $i$'s utility, condition C5 says to take that one in which all links are maintained.
Hence, the myopic equilibrium is unique for the incomplete information setting.
In addition to the above, in the complete information setting condition C3 uniquely determines the action $s_{\tilde{i} \tilde{j}}$ in case bacterium $\tilde{i}$ approaches bacterium $\tilde{j}$. 
\end{IEEEproof}

\begin{rem}
The uniqueness of the equilibrium results from the fact that the 
best action of each bacterium is independent from the actions of the other bacteria. 
For instance, if $i$ has the incentive to maintain (form) a link with $j$, its best action is $a_{ij}^{(n)}=1$ ($s_{ij}^{(n)}=1$), regardless of $j$'s action. 
Indeed, if $j$ decides to maintain (form) the link then $i$ achieves its goal, whereas if $j$ decides not to maintain (form) the link then $i$ does not incur any additional cost with respect to the action $a_{ij}^{(n)}=0$ ($s_{ij}^{(n)}=0$).
This property implies that the resulting myopic equilibrium is robust to changes in the actions of the other bacteria.
\end{rem}

\begin{rem}
Proposition \ref{prop:uniq} implies that in each time slot $n$, given the network $\mathcal{G}^{(n)}$ and the parameters of the system, the decisions of the bacteria are unique.
However, the evolution of the network $\mathcal{G}^{(n)}$ is not unique, because it depends both on these decisions 
and on the randomness of the meetings among bacteria. 
Throughout the paper we implicitly assume that bacteria, in each time slot $n$, adopt the unique myopic equilibrium, and we study the possible evolutions of the network $\mathcal{G}^{(n)}$ depending of the parameters of the system. 
\end{rem}

\vspace{-0.2cm}
\subsection{micro-colonies and stable networks} 
\label{sec:micro} 


We say that the link $(i,j) \in \mathcal{G}^{(n)}$ is \emph{stable} if $(i,j) \in \mathcal{G}^{(m)}$, $\forall \, m > n$, regardless of the realization of the meetings among bacteria; whereas, if there is a positive probability that $(i,j) \notin \mathcal{G}^{(m)}$ for some $m>n$, we say that the link is \emph{unstable}.

\vspace{-0.1cm}
\begin{defn}[micro-colony]
$\mathcal{M}$ 
is a micro-colony in time instant $n$ if and only if the following conditions are satisfied
\begin{itemize}
\item[A1)] $\mathcal{M}$ is a subset of a component;
\item[A2)] $\forall \, i,j \in \mathcal{M}$ such that $(i,j) \in \mathcal{G}^{(n)}$, we have that $(i,j)$ is stable;
\item[A3)] $\forall \, i \in \mathcal{M}$ and $j \notin \mathcal{M}$ such that $(i,j) \in \mathcal{G}^{(n)}$, we have that $(i,j)$ is unstable;
\item[A4)] $\abs{\mathcal{M}} \geq 2$.
\end{itemize}
\end{defn}

A1 states that all the bacteria belonging to the micro-colony are connected to each others.
A2 is a stability condition, it states that the links among bacteria in the micro-colony are never broken (hence, the size of a micro-colony can only grow in time), whereas by A3 a bacterium belonging to the same component $\mathcal{C}$ of a micro-colony, but not belonging to the micro-colony, may leave the component in the future.
A4 excludes a singleton bacterium from being considered a trivial form of micro-colony.

Exploiting the definition of micro-colony, we can now define a network stability concept.

\vspace{-0.1cm}
\begin{defn}[Stable-network]
$\mathcal{G}^{(n)}$ is a stable network if each bacterium belongs to a micro-colony.
\end{defn}

As a consequence, if $\mathcal{G}^{(n)}$ is a stable network then no link will ever be formed or broken 
and each bacterium is linked to at least another bacterium (i.e., no bacterium moves freely along the surface).
A stable network can be interpreted as an important step toward the formation of a biofilm. 
Indeed, if there are many bacteria that never link to micro-colonies then biofilm formation is greatly reduced and the integrity of resulted biomass is severely degraded \cite{Toole00}.

Let $\mathcal{G}^S$ be the set of stable networks. 
We say that $\mathcal{G}^{(n)}$ \emph{converges to a stable network} if and only if $\lim_{n \to +\infty} \mathcal{G}^{(n)}$ exists and belongs to $\mathcal{G}^S$.
We say that $\mathcal{G}^{(n)}$ \emph{does not converge to a stable network} if and only if $\lim_{n \to +\infty} \mathcal{G}^{(n)}$ does not exists, or it exists but it does not belong to $\mathcal{G}^S$. 
Notice that the evolution of the network $\mathcal{G}^{(n)}$ is a random process due to the randomness of the meetings among bacteria. 
However, for some set of parameters it may be the case that $\mathcal{G}^{(n)}$ converges (does not converges) to a stable network \emph{with probability} $1$, regardless of the evolution it follows. 
Formally, denoting by $P[A]$ the probability of a generic event $A$, we say that $\mathcal{G}^{(n)}$ \emph{converges to a stable network with probability} $1$ if and only if $P \left[ \lim_{n \to \infty} \mathcal{G}^{(n)} \in \mathcal{G}^S \right] = 1$, $\mathcal{G}^{(n)}$ \emph{does converges to a stable network with probability $1$} if and only if $P \left[ \lim_{n \to \infty} \mathcal{G}^{(n)} \notin \mathcal{G}^S \right] = 1$. 

\begin{rem}
``$\mathcal{G}^{(n)}$ does not converge with probability $1$'' is not the complementary of ``$\mathcal{G}^{(n)}$ converges with probability $1$''; if fact, it is possible that the probability that a network converges is positive but lower than $1$.
In these cases the convergence is determined by chance, by the realization of the meeting among bacteria.
\end{rem} 



\section{Evolution To A Stable Network} 
\label{sec:ana} 

In this section we analyze the dynamic network formation game formally described in Section \ref{sec:dyn}, in both the complete and the incomplete information settings.
As a main contribution of this section we analytically derive the conditions on the system parameters under which the network $\mathcal{G}^{(n)} $ converges to a stable network with probability $1$, under which $\mathcal{G}^{(n)}$ does not converge to a stable network with probability $1$, and under which the convergence of $\mathcal{G}^{(n)}$ is determined by chance.

\subsection{Complete information setting} 

In this subsection we analyze the complete information settings. 
In the first result, we characterize the actions of low type bacteria, the action of high type bacteria when linked to some other bacteria, and the actions of singleton high type bacteria that meet other singleton bacteria.

\begin{lemma}
For each bacterium $i$ and for each neighbor $j \in \mathcal{N}_i^{(n)}$ we have $a_{ij}^{(n)} = 1$.

For each singleton low type bacterium $\tilde{i}$ that approaches a singleton bacterium $\tilde{j}$ in time instant $n$, we have $s_{\tilde{i}\tilde{j}}^{(n)}=1$.

For each singleton high type bacterium $\tilde{i}$ that approaches a singleton low type bacterium $\tilde{j}$ in time instant $n$, we have $s_{\tilde{i}\tilde{j}}^{(n)}=1$ if and only if $f(L) \geq c(H)$.

For each singleton high type bacterium $\tilde{i}$ that approaches a singleton high type bacterium $\tilde{j}$ in time instant $n$, we have $s_{\tilde{i}\tilde{j}}^{(n)}=1$ if and only if $f(H) \geq c(H)$.
\label{lem:com_act}
\end{lemma}

\vspace{-0.1cm}
\begin{IEEEproof}
Given the utility structure (\ref{eq:ut}), it is trivial that 1) because $c(L)=0$, low type bacteria always benefit from being linked to other bacteria, 2) a  singleton high type bacterium increases its utility if it links with a singleton low type bacterium if and only if $f(L) \geq c(H)$, and 3) a  singleton high type bacterium increases its utility if it links with a singleton high type bacterium if and only if $f(H) \geq c(H)$.
It remains to show that high type bacteria always want to maintain a link. 
First, notice that a bacterium does not break a strict subset of its links because the cost to pay for a single link is equal to the cost to pay for multiple links.
Hence, a bacterium selects either to break all links or to maintain all links.
To conclude our proof we show that the utility of each bacterium $i$ is non decreasing in time, which implies that no bacterium has the incentive to break all of its links returning to the initial situation. 
Let $u_i^{(n)}$ the utility of $i$ at time slot $n$.
Assume $i$ is singleton at the beginning of the $n$-th slot, if it does not form a link during the $n$-th slot then $u_i^{(n+1)}=u_i^{(n)}$, whereas if it forms a link then $u_i^{(n+1)} \geq u_i^{(n)}$.
Assume $i$ is non singleton at the beginning of the $n$-th slot, if no bacterium links to its component during the $n$-th slot then $u_i^{(n+1)}=u_i^{(n)}$, whereas if a bacterium links to its component then $u_i^{(n+1)} > u_i^{(n)}$. 
\end{IEEEproof}

Lemma \ref{lem:com_act} shows that, in the complete information setting, all the links are stable.
In fact, when forming a link, a bacterium knows in advance it will increase its utility at the end of the current time slot, and such increment can only increase in time because new bacteria can join the component it belongs to.
As a consequence, since bacteria do not break links and do not leave the components they belong to, the concept of micro-colony coincides with the concept of component with size at least  $2$. 
This is remarked in the following proposition. 

\vspace{-0.1cm}
\begin{proposition}
All components $\mathcal{C}$ of size at least $2$ are micro-colonies. 
\label{prop:com_stable}
\end{proposition}

\vspace{-0.1cm}
\begin{IEEEproof}
Lemma \ref{lem:com_act} proves that all components are stable, hence all components of size at least $2$ satisfy A1-A4.
\end{IEEEproof}

Another implication of Lemma \ref{lem:com_act} is that singleton high type bacteria do not form links with singleton bacteria if $f(H) < c(H)$.
In this case, high type bacteria must wait for low type bacteria to form micro-colonies before starting to form links.
One may wonder if there exists a minimum size a micro-colony of low type bacteria must have before a high type bacterium joins it. 
Proposition \ref{prop:com_low_bound} answers positively to this question.

\begin{proposition}
Let $\mathcal{M}$ be a micro-colony. 
If $f(H) < c(H)$ and $\abs{\mathcal{M}} < N_{th,1} \triangleq \frac{c(H)-f(L)}{\delta f(L)}+2$ then $\mathcal{M}$ does not contain bacteria of high type.
\label{prop:com_low_bound}
\end{proposition}


\begin{IEEEproof}
We prove the statement by contradiction. 
Assume $\mathcal{M}$ is a micro-colony that contains high type bacteria and $\abs{\mathcal{M}} < N_{th,1}$.
Denote by $i$ the first high type bacterium that linked to the micro-colony, by $n$ the time slot in which this happened, and by $\underline{\mathcal{M}}$ the resulting micro-colony after the link formation.
Notice that the condition $f(H) < c(H)$ excludes the possibility that the micro-colony generated from a link between two high type bacteria, hence $\underline{\mathcal{M}}$ is formed by low type bacteria except for $i$.
Then
\ba
u_i \left( \mathbf{t}, \mathcal{G}^{(n)} \right) \leq 
f(L) + \left( \abs{\underline{\mathcal{M}}} - 2 \right) \delta f(L) - c(H) \leq 
f(L) + \left( \abs{\mathcal{M}} - 2 \right) \delta f(L) - c(H)  < 0 , \nonumber
\end{align}
where the first inequality is valid because $i$'s maximum utility is achieved when it links to a bacterium that is directly linked with all the other bacteria, the second inequality is valid because a micro-colony can only increase in size, and the third inequality is valid because $\abs{\mathcal{M}} < N_{th,1}$.
This contradicts the statement that $i$'s utility in non decreasing in time (see proof Lemma \ref{lem:com_act}).
\end{IEEEproof}

\begin{rem}
$N_{th,1}-1$ represents the minimum size a micro-colony of low type bacteria must have such that a high type bacterium may have an incentive to join the micro-colony. 
\end{rem} 

$\abs{\mathcal{M}} \geq N_{th,1}-1$ is a necessary condition such that a high type bacterium joins the micro-colony $\mathcal{M}$.
Now we investigate the existence of a sufficient condition, i.e., the existence of a minimum size $\abs{\mathcal{M}}$ that guarantees that high type bacteria always want to link to a bacterium belonging to $\mathcal{M}$.
We consider only the case $f(L) < c(H)$, because Lemma \ref{lem:com_act} guarantees that for $f(L) \geq c(H)$ high type bacteria always want to form a link.

\begin{proposition}
If $f(L) < c(H)$ and $\delta > \frac{c(H) - f(L)}{c(H)}$, then there exists $N_{th,2} (f(L) , c(H) , \delta)$, increasing in $c(H)$ and decreasing in $f(L)$ and $\delta$, such that if a high type bacterium $\tilde{i}$ meets a bacterium $\tilde{j}$ belonging to a micro-colony $\mathcal{M}$ with size $\abs{\mathcal{M}} \geq N_{th,2} (f(L) , c(H) , \delta)$, then $s_{\tilde{i}\tilde{j}}^{(n)}=1$. 
\label{prop:com_upp_bound}
\end{proposition}

\begin{IEEEproof}
The utility $\tilde{i}$ obtains forming a link with $\tilde{j}$ is
$
u_{\tilde{i}} \geq
\sum_{j=0}^{\abs{\mathcal{M}}-1} \delta^j f(L) - c(H) = 
f(L) \frac{1 - \delta^{\abs{\mathcal{M}}}}{1 - \delta} - c(H),
$
where the first inequality is valid because $\tilde{i}$'s lowest utility is achieved when all bacteria in $\mathcal{M}$ (except for $\tilde{i}$) have low type, when they are aligned, and $\tilde{j}$ is located in one extreme of the line.
Hence, $u_{\tilde{i}} \geq 0$ (i.e., $\tilde{i}$ wants to form a link) if
\ba
f(L) \dfrac{1 - \delta^{\abs{\mathcal{M}}}}{1 - \delta} \geq c(H)
\label{eq:bah}
\end{align}
For $\abs{\mathcal{M}} = 1$ the inequality (\ref{eq:bah}) is not satisfied because $f(L) < c(H)$.
Since the left side of (\ref{eq:bah}) increases in $\abs{\mathcal{M}}$ and since for $\abs{\mathcal{M}} \to +\infty$ the inequality (\ref{eq:bah}) holds strictly (because $\delta > \frac{c(H) - f(L)}{c(H)}$), then there exists finite $N_{th,2} (f(L) , c(H) , \delta)$ such that for $\abs{\mathcal{M}} = N_{th,2} (f(L) , c(H) , \delta)$ the inequality (\ref{eq:bah}) holds with equality.
Hence, for $\abs{\mathcal{M}} \geq N_{th,2} (f(L) , c(H) , \delta)$ the inequality \ref{eq:bah} holds.
Finally, since the left hand side of (\ref{eq:bah}) increases in $f(L)$ and $\delta$, whereas the right hand side of (\ref{eq:bah}) increases in $c(H)$, we have that $N_{th,2} (f(L) , c(H) , \delta)$ increases in $c(H)$ and decreases in $f(L)$ and $\delta$.
\end{IEEEproof}

\begin{rem}
$N_{th,2}$ represents a \emph{critical size} for a micro-colony $\mathcal{M}$, above which high type bacteria always want to link to a bacterium belonging to $\mathcal{M}$.
\end{rem} 

Proposition \ref{prop:com_upp_bound} implies that high type bacteria cannot remain singleton forever if there exists a micro-colony with size at least $N_{th,2}$, because they would eventually be attracted by the micro-colony.
This allows us to characterize the structure of a network that does not converge to a stable network.

\begin{theorem}
If $\mathcal{G}^{(n)}$ does not converge to a stable network, then $\mathcal{G}^{(n)}$ converges to a network in which all high type bacteria are singleton and each low type bacterium belongs to a micro-colony with size lower than $N_{th,2}$.
\label{teo:com_sing}
\end{theorem}

\begin{IEEEproof}
We prove that all high type bacteria will be singleton by contradiction. 
Assume the high type bacterium $i$ belongs to a micro-colony in a generic time instant $n$ and let $(i, j)$ the first link formed by $i$.
Since a micro-colony can only grow in size, the utility a singleton high type bacterium obtains by forming a link with $j$ after time instant $n$ is at least as high as the utility $i$ achieved when it formed the link with $j$, and as a consequence it is higher than the utility it obtains being singleton. 
Hence, a high type bacterium cannot stay singleton forever because, if it does not join a micro-colony in the meantime, it will eventually meet $j$ (such event happens with positive probability in each time slot) and form a stable link with it.
This contradicts the fact that $\mathcal{G}^{(n)}$ does not converge to a stable network.

A low type bacterium cannot stay singleton forever because, if it does not join a micro-colony in the meantime, it will eventually meet another low type bacterium (such event happens with positive probability in each time slot) and form a stable link with it.
Finally, if a micro-colony of low type bacteria has a size larger than $N_{th,2}$, then a high type bacterium will eventually meet a bacterium belonging to the micro-colony and, because of Proposition \ref{prop:com_upp_bound}, it will join the micro-colony, contradicting the fact that all high type bacteria are singleton.
\end{IEEEproof}

Now we analytically derive the conditions on the system parameters under which the network $\mathcal{G}^{(n)} $ converges to a stable network with probability $1$, under which $\mathcal{G}^{(n)}$ does not converge to a stable network with probability $1$, and under which the convergence of $\mathcal{G}^{(n)}$ is determined by chance.

\begin{theorem}
$\mathcal{G}^{(n)}$ converges with probability $1$ to a stable network if and only if either 1) $f(H) \geq c(H)$, or 2) $(1 + \delta) f(L) \geq c(H)$ and $K(L) \geq 2$, or 3) $K(H) = 0$.

$\mathcal{G}^{(n)}$ does not converge to a stable network with probability $1$ if and only if $f(H) < c(H)$, $K(L) < N_{th,1}-1$, and $K(H) > 0$.

\label{teo:com_conv}
\end{theorem}

\begin{IEEEproof}
If $f(H) \geq c(H)$ then a high type bacterium cannot be singleton forever, because it would eventually meet another high type bacterium (singleton or not) and form a stable link with it.
Hence, exploiting Theorem \ref{teo:com_sing}, we have $f(H) \geq c(H)$ implies that $\mathcal{G}^{(n)}$ converges with probability $1$ to a stable network.

If $(1 + \delta) f(L) \geq c(H)$ and $K(L) \geq 2$ then a high type bacterium cannot be singleton forever, because it would eventually meet a low type bacterium that is connected to at least another low type bacterium and form a stable link with it.
Hence, exploiting Theorem \ref{teo:com_sing}, we have $(1 + \delta) f(L) \geq c(H)$ implies $\mathcal{G}^{(n)}$ converges with probability $1$ to a stable network.

If $K(H) = 0$, then there are no high type bacteria and a network of low type bacteria always converges with probability $1$ to a stable network.

If $f(H) < c(H)$, $K(L) < N_{th,1}-1$, and $K(H) > 0$, then the largest colony of low type bacteria has a size lower than $N_{th,1}-1$.
Hence, for Theorem \ref{teo:com_sing} we have $f(H) < c(H)$ and $K(L) < N_{th,1}-1$ implies that $\mathcal{G}^{(n)}$ does not converge to a stable network with probability $1$. 

Finally, we prove that if $f(H) < c(H)$, $(1 + \delta) f(L) < c(H)$, $K(L) \geq N_{th,1}-1$, and $K(H) > 0$, then the probability that $\mathcal{G}^{(n)}$ converges to a stable network is positive but lower than $1$. 
On one hand, assume that there exists a time instant $n$ such that no low type bacterium is singleton and the size of all micro-colonies is lower than $N_{th,1} - 1$ (e.g., low type bacteria are connected in couples).
Notice that this event happens with positive probability.
Proposition \ref{prop:com_low_bound} guarantees that no high type bacterium will ever link to a micro-colony, and the network does not converge to a stable network. 
On the other hand, assume that there exists a time instant $n$ such that at least $N_{th,1}-1$ low type bacteria are connected together in a star topology. 
Notice that this event happens with positive probability.
A high type bacteria $i$ will eventually meet the low type bacteria located at the center of the star topology.
Denote by $u_i$ the utility $i$ would obtain if it forms the link, we have
$u_i \geq  
f(L) + \left( N_{th,1} - 2 \right) \delta f(L) - c(H) = 0$; 
hence, $i$ will form a stable link and Theorem \ref{teo:com_sing} guarantees that $\mathcal{G}^{(n)}$ converges to a stable network. 
\end{IEEEproof}

\begin{rem}
Theorem \ref{teo:com_conv} implies that if $f(H) < c(H)$, $(1 + \delta) f(L) < c(H)$, and $K(L) \geq N_{th,1}-1$, then the convergence of $\mathcal{G}^{(n)}$ is determined by chance.
\end{rem} 

\begin{rem}
If $f(H) < c(H)$, low type bacteria play a fundamental role in the initial phase of the micro-colonies formation process. 
In fact, in this case high type bacteria
have an incentive to link with other bacteria only after low type bacteria have formed micro-colonies of a large enough size. 
\end{rem} 

To understand why and how $\mathcal{G}^{(n)}$ converges (does not converges) to a stable network, in the following we describe some examples. 
If $f(H) \geq c(H)$ then two singleton high type bacteria form a link when they meet, and this enables the formation of a micro-colony, as shown by Fig. \ref{fig:test}a.
If $(1 + \delta) f(L) \geq c(H)$ then a high type bacterium meeting a pair of low type bacteria forms a link with them, and this enables the formation of a micro-colony, as shown by Fig. \ref{fig:test}b.
If $f(H) < c(H)$, $K(L) < N_{th,1}$ and $K(H) > 0$, then the network cannot evolve in a condition such that a high type bacterium has the incentive to form a link, as shown by Fig. \ref{fig:test}c.
If $f(H) < c(H)$, $(1 + \delta)f(L) < c(H)$, $K(L) < N_{th,1}$, and $K(H) > 0$, then it is possible that enough low type bacteria connect together giving an incentive for high type bacteria to join the micro-colony, but it is also possible that low type bacteria form many small size micro-colonies that high type bacteria are not willing to join, both possibilities are shown in Fig. \ref{fig:test}d.

\begin{figure}
     \centering
          \includegraphics[width=\figwww]{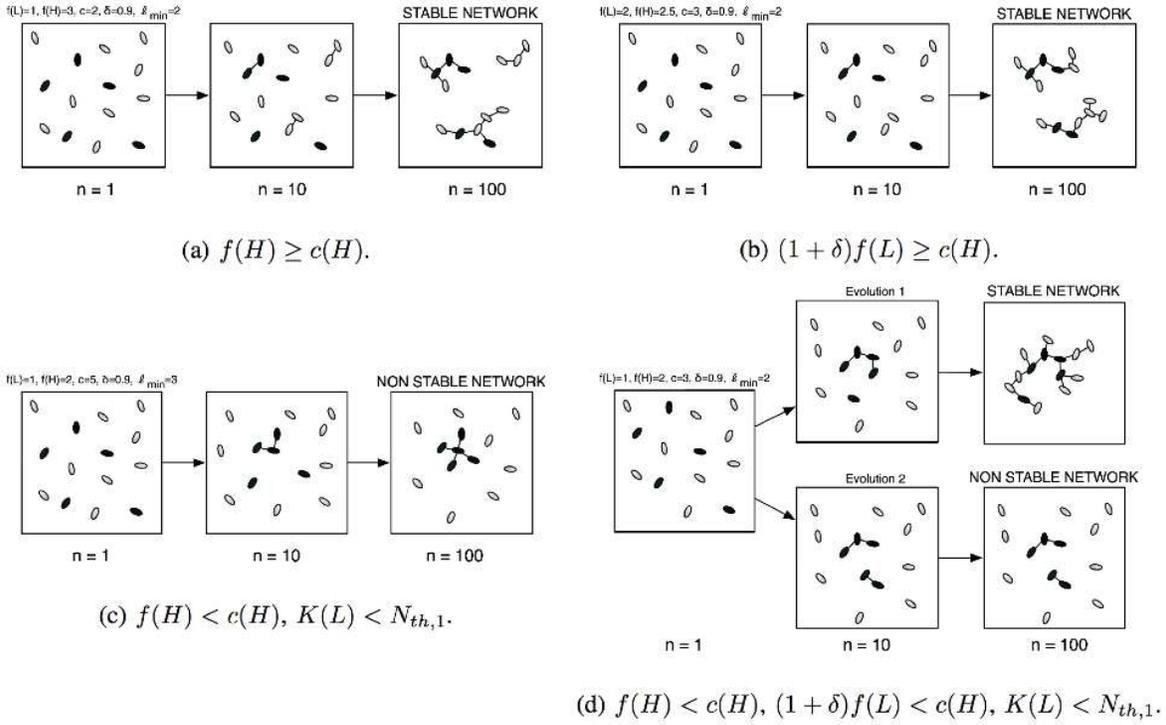}
\caption{Possible evolutions of $\mathcal{G}^{(n)}$ in the complete information setting. Black and white cells refer to low and high type bacteria, respectively.}
\label{fig:test}
\end{figure}

Now we investigate the \emph{robustness} of a stable network, that is, we wonder what happens to a stable network if some links are broken regardless from the actions chosen by bacteria (e.g., some bacteria can die). 
Does the network converges again to a stable network?
Theorem \ref{teo:com_rob} characterizes the conditions on the system parameters and on the number of broken links such that the answer to the above question is positive. 

\begin{theorem}
Let $\mathcal{G}^{(n)}$ a stable network, and assume that $L$ links are removed from it.
If either $f(H) \geq c(H)$ or $(1 + \delta) f(L) \geq c(H)$, then $\mathcal{G}^{(n)}$ converges again to a stable network, regardless from the number $L$ of removed links.  
If $f(H) < c(H)$, $(1 + \delta) f(L) < c(H)$, and $K(L) \geq N_{th,1}-1$, then
$\mathcal{G}^{(n)}$ converges again to a stable network if $L \leq \frac{K(H)}{2 (N_{th,2}+1)}$.
\label{teo:com_rob}
\end{theorem}

\begin{IEEEproof}
If $f(H) \geq c(H)$ or $(1 + \delta) f(L) \geq c(H)$, then using the same arguments as in the proof of Theorem \ref{teo:com_conv} $\mathcal{G}^{(n)}$ converges again to a stable network.

Now assume $f(H) < c(H)$, $(1 + \delta)f(L) < c(H)$, and $K(L) < N_{th,1}$. 
Denote by $M$ the number of micro-colonies in $\mathcal{G}^{(n)}$ containing at least one high type bacterium, by $K(m)$ the number of high type bacteria belonging to the micro-colony $m$, and by $L(m)$ the number of links broken in micro-colony $m$. 
We have $\sum_{m=1}^M K(m) = K(H)$ and $\sum_{m=1}^M L(m) = L$.
If $L(m) < 1$ for some $m$, then the micro-colony $m$ is still present in the new network, and since $m$ contains high type bacteria, for Theorem \ref{teo:com_sing}, $\mathcal{G}^{(n)}$ will converge again to a stable network. 
If $K(m) \geq N_{th,2}+1$ and  $L(m) < \lfloor \frac{K(m)}{N_{th,2}+1} \rfloor $, where $\lfloor \cdot \rfloor$ is the largest integer smaller than the argument, then the micro-colony $m$ is divided into $L(m)$ component, and at least one of these components has a size equal to or larger than $N_{th,2}+1$. 
Hence, for Proposition \ref{prop:com_upp_bound}, such component is a micro-colony, and since it contains high type bacteria, for Theorem \ref{teo:com_sing}, $\mathcal{G}^{(n)}$ will converge again to a stable network. 
As a consequence, the minimum number of links to break such that there might be a possibility that $\mathcal{G}^{(n)}$ does not converge to a stable network is 
\ba
L_{min} \triangleq \sum_{m=1}^M \max \{ 1, \floor*{ \dfrac{K(m)}{N_{th,2}+1} }  \} > \sum_{m=1}^M \dfrac{K(m)}{2 (N_{th,2}+1)} = \dfrac{K(H)}{2 (N_{th,2}+1)} \;\; ,
\nonumber
\end{align}
where the inequality is valid because if $\frac{K(m)}{N_{th,2}+1} > 1$ then $ \max \{ 1, \floor*{ \frac{K(m)}{N_{th,2}+1} }  \} = \floor*{ \frac{K(m)}{N_{th,2}+1} } > \frac{K(m)}{2 (N_{th,2}+1)} $, whereas if $\frac{K(m)}{2(N_{th,2}+1)} \leq 1$ then $\max \{ 1, \floor*{ \frac{K(m)}{N_{th,2}+1} }  \}  = 1 > \floor*{ \frac{K(m)}{2(N_{th,2}+1)} }$.
\end{IEEEproof}

\begin{rem}
If the number of high type bacteria $K(H)$ is large compare to the critical size $N_{th,2}$, then the number of links to break to have a possibility that $\mathcal{G}^{(n)}$ does not converge again to a stable network is large.
\end{rem}

\subsection{Incomplete information setting} 

In this subsection we analyze the incomplete information settings. 

In the first result, we shows that low type bacteria never break links, as a consequence a pair of low type bacteria is a micro-colony. 
Also, we characterize the conditions under which a pair of high and low type bacteria and a pair of high type bacteria are micro-colonies.

\begin{lemma}
For each low type bacterium $i$ and for each neighbor $j \in \mathcal{N}_i^{(n)}$ we have $a_{ij}^{(n)} = 1$.

A pair of high and low type bacteria is a micro-colony if and only if $f(L) \geq c(H)$.

A pair of high type bacteria is a micro-colony if and only if $f(H) \geq c(H)$.
\label{lem:inc_act}
\end{lemma}

\begin{IEEEproof}
Because $c(L)=0$, low type bacteria always benefit from being linked to other bacteria, hence they always want to maintain a link. 
A high type bacterium $i$ obtains a higher utility being singleton than being paired with a low (high) type bacterium $j$ if and only if $f(L) < c(H)$ ($f(H) < c(H)$).
In this case, since there is a positive probability that no additional bacterium forms a link with $i$ or $j$ before $\ell_{ij}^{(n)} > \ell_{min}$, $i$ may eventually break the link $(i,j)$.
\end{IEEEproof}

\begin{rem}
In the incomplete information setting low type bacteria never break links but, differently from the complete information setting, high type bacteria can break links. 
As a consequence not all the components are micro-colonies.
\end{rem}

An implication of Lemma \ref{lem:inc_act} is that singleton high type bacteria form unstable links with other singleton bacteria if $f(H) < c(H)$.
In the next proposition we generalize such a results, and show that all links between a high type bacterium and another bacterium are unstable if the size of the component they belong to is below a certain threshold. 

\begin{proposition}
Let $\mathcal{M}$ be a micro-colony. 
If $f(H) < c(H)$ and $\abs{\mathcal{M}} < N_{th,3} \triangleq \frac{c(H)}{f(H)}+1$ then $\mathcal{M}$ does not contain bacteria of high type.
\label{prop:inc_low_bound}
\end{proposition}

\begin{IEEEproof}
We prove the statement by contradiction. 
Assume $\mathcal{M}$ is a micro-colony that contains high type bacteria and $\abs{\mathcal{M}} < N_{th,3}$.
Denote by $i$ the first high type bacterium that linked to the micro-colony, by $n$ the time slot in which this happened, and by $\underline{\mathcal{M}}$ the resulting micro-colony after the link formation.
Notice that the condition $f(H) < c(H)$ excludes the possibility that the micro-colony generated from a link between two high type bacteria, hence $\underline{\mathcal{M}}$ is formed by low type bacteria except for $i$.
Then 

$u_i \left( \mathbf{t}, \mathcal{G}^{(n)} \right) \leq 
\left( \abs{\underline{\mathcal{M}}} - 1 \right) f(L) - c(H) \leq
\left( \abs{\mathcal{M}} - 1 \right) f(L) - c(H)  < 0, 
$

where the first inequality is valid because $i$'s maximum utility is achieved when all bacteria in $\underline{\mathcal{M}}$ are directly linked with $i$, the second inequality is valid because a micro-colony can only increase in size, and the third inequality is valid because $\abs{\mathcal{M}} < N_{th,3}$.
Since there is a positive probability that no additional bacterium joins the micro-colony $\mathcal{M}$ before the lengths of $i$'s links are larger than $\ell_{min}$, $i$ may eventually leave $\mathcal{M}$.
This implies that $\mathcal{M}$ is not a micro-colony.
\end{IEEEproof}

\begin{rem}
We have $N_{th,3} < N_{th,1}$.
Indeed, in the incomplete information setting a high type bacterium $i$ can form an unstable link with another high type bacterium, and before $i$ can break this link other high type bacteria can form direct links with $i$. 
Since $i$ can be directly connected with many high type bacteria before having the possibility to break the original link, in the incomplete information setting it is possible to form smaller micro-colonies containing high type  bacteria than in the complete information setting. 
Finally, $N_{th,3}$ does not depends on $\delta$ because of this possibility of creating many direct links.
\end{rem}

Now we investigate the existence of a minimum size of a component $\mathcal{C}$ that guarantees that all the links are stable, i.e., such that $\mathcal{C}$ is a micro-colony.
We consider only the case $f(L) < c(H)$, because Lemma \ref{lem:inc_act} proves that for $f(L) \geq c(H)$ all links are stable.

\begin{proposition}
If $f(L) < c(H)$ and $\delta > \frac{c(H) - f(L)}{c(H)}$, then there exists $N_{th,4} (f(L) , c(H) , \delta)$, increasing in $c(H)$ and decreasing in $f(L)$ and $\delta$, such that all the components $\mathcal{C}$ with size $\abs{\mathcal{C}} \geq N_{th,4} (f(L) , c(H) , \delta) + 1$ are micro-colonies.
\label{prop:inc_upp_bound}
\end{proposition}

\begin{IEEEproof}
If $f(L) < c(H)$ and $\delta > \frac{c(H) - f(L)}{c(H)}$ then the utility of a high type bacterium $i$ belonging to a component  $\mathcal{C}$ satisfies

$u_i \left( \mathbf{t}, \mathcal{G}^{(n)} \right) \geq
\sum_{j=0}^{\abs{\mathcal{C}}-2} \delta^j f(L) - c(H) = 
f(L) \dfrac{1 - \delta^{\abs{\mathcal{C}}-1}}{1 - \delta} - c(H), 
$

where the first inequality is valid because $i$'s lowest utility is achieved when all bacteria in $\mathcal{M}$ (except for $i$) have low type, when they are aligned, and $i$ is located in one extreme of the line.
Hence, $u_i \left( \mathbf{t}, \mathcal{G}^{(n)} \right) \geq 0$, meaning that $i$ does not break any link, if
\ba
f(L) \dfrac{1 - \delta^{\abs{\mathcal{C}}-1}}{1 - \delta} \geq c(H)
\label{eq:bah2}
\end{align}
For $\abs{\mathcal{C}} = 2$ inequality (\ref{eq:bah2}) is not satisfied because $f(L) < c(H)$.
For $\abs{\mathcal{C}} \to +\infty$ inequality (\ref{eq:bah2}) holds strictly because $\delta > \frac{c(H) - f(L)}{c(H)}$.
The left side of (\ref{eq:bah2}) increases in $\abs{\mathcal{C}}$, hence there exists $N_{th,4} (f(L) , c(H) , \delta)$ such that for $\abs{\mathcal{C}} \geq N_{th,4} (f(L) , c(H) , \delta)$ inequality \ref{eq:bah2} holds.
Finally, since the left hand side of (\ref{eq:bah2}) increases in $f(L)$ and $\delta$, whereas the right hand side of (\ref{eq:bah2}) increases in $c(H)$, we have that $N_{th,4} (f(L) , c(H) , \delta)$ increases in $c(H)$ and decreases in $f(L)$ and $\delta$.
\end{IEEEproof}

\begin{rem}
$N_{th,4}$ represents a \emph{critical size} for a component $\mathcal{C}$, above which high type bacteria never leave the component. 
\end{rem}

Proposition \ref{prop:inc_upp_bound} implies that high type bacteria cannot remain singleton forever if there exists a micro-colony with size at least $N_{th,4}-\ell_{min}+1$, because it eventually happens that $\ell_{min}-1$ bacteria link sequentially to the micro-colony, forming a micro-colony of size at least $N_{th,4}$ that would eventually attract all the singleton bacteria.
This allows us to characterize the structure of a network that does not converge to a stable network.

\begin{theorem}
If $\mathcal{G}^{(n)}$ does not converge to a stable network, then $\mathcal{G}^{(n)}$ converges to a network in which no high type bacterium belongs to a micro-colony and each low type bacterium belongs to a micro-colony with size lower than $N_{th,4}-\ell_{min}+1$.
\label{teo:inc_sing}
\end{theorem}

\begin{IEEEproof}
We prove that no high type bacterium will be part of some micro-colonies by contradiction. 
Assume the high type bacterium $i$ belongs to a micro-colony in a generic time instant $n$ and let $(i, j)$ the first link formed by $i$.
Since a micro-colony can only grow in size, the utility a singleton high type bacterium obtains by forming a link with $j$ after time instant $n$ is at least as high as the utility $i$ achieved when it formed the link with $j$, and as a consequence it is higher than the utility it obtains being singleton. 
Hence, each singleton high type bacterium, if it does not join a micro-colony in the meantime, will eventually meet $j$ (such event happens with positive probability in each time slot) and form a stable link with it.
This contradicts the fact that $\mathcal{G}^{(n)}$ does not converge to a stable network.
If a low type bacterium does not belong to a micro-colony, it will eventually meet a micro-colony or another low type bacterium (such event happens with positive probability in each time slot) and form a stable link.
Finally, if a micro-colony of low type bacteria has a size larger than $N_{th,4}-\ell_{min}$, then it will eventually happen that $\ell_{min}$ high type bacteria link to that colony in subsequent time slots. 
Proposition \ref{prop:inc_upp_bound} guarantees that the new component is a micro-colony, contradicting the fact that no high type bacterium will be part of some micro-colonies.
\end{IEEEproof}

Now we analytically derive the conditions on the system parameters under which the network $\mathcal{G}^{(n)} $ converges to a stable network with probability $1$, under which $\mathcal{G}^{(n)}$ does not converge to a stable network with probability $1$, and under which the convergence of $\mathcal{G}^{(n)}$ is determined by chance.

\begin{theorem}
Consider the case $\ell_{min} = 1$.
Then $\mathcal{G}^{(n)}$ converges with probability $1$ to a stable network if and only if either 1) $f(H) \geq c(H)$, or 2) $(1 + \delta) f(L) \geq c(H)$ and $K(L) \geq 2$, or 3) $K(H) = 0$; whereas $\mathcal{G}^{(n)}$ does not converge to a stable network with probability $1$ if and only if $f(H) < c(H)$, $K(L) < N_{th,1}-1$, and $K(H) > 0$. 

Consider the case $\ell_{min} \geq 2$. 
Then $\mathcal{G}^{(n)}$ converges with probability $1$ to a stable network if $K(H) \geq N_{th,4}-2$ and $K \geq N_{th,4}$; whereas $\mathcal{G}^{(n)}$ does not converge to a stable network with probability $1$ if $f(H) < c(H)$, $K < N_{th,3}$, and $K(H)>0$.
\label{teo:inc_conv}
\end{theorem}

\begin{IEEEproof}
If $\ell_{min} = 1$, then the evolution of the network $\mathcal{G}^{(n)}$ is as in the complete information case, because a bacterium that forms a link can break it immediately (at the beginning of the next slot), before other bacteria can join the component it belongs to.
Hence, Theorem \ref{teo:com_conv} holds. 

Now consider the case $\ell_{min} \geq 2$.
Assume a high type $i$ links with a low type $j$ in time instant $n$.
Even though this link is unstable, a new bacteria $k$ can link with $i$ in time instant $m$, such that $n < m < n + \ell_{min}$.
This means that $i$ does not break the link in time instant $n + \ell_{min}$ (i.e., when $\ell_{ij}^{(n)} < \ell_{min}$), because it is never convenient to break only parts of the links, and the link $(i,k)$ cannot be broken.
If we iterate this reasoning, a component of whatever side can be formed before a bacteria has the possibility to break its links. 
Hence, if $K(H) \geq N_{th,4}-2$ and $K \geq N_{th,4}$, high type bacteria will eventually belong to micro-colonies, because they can eventually form a component of size at least $N_{th,4}$ and Proposition \ref{prop:inc_upp_bound} guarantees that this is a micro-colony. 
On the other hand, if there are high type bacteria in the system (i.e., $K(H)>0$), they cannot benefit from linking together (i.e., $f(H) < c(H)$), and the total number of bacteria do not allow to form a component of size at least $N_{th,3}$, then Proposition \ref{prop:inc_low_bound} guarantees that $\mathcal{G}^{(n)}$ does not converge to a stable network with probability $1$. 
\end{IEEEproof}

\begin{rem}
If the minimum link length is $1$, then complete and incomplete information converge under the same conditions.
\end{rem} 

\begin{rem}
If the minimum link length is larger than $1$, then in the incomplete information settings bacteria can form large size components before they have the possibility to break their links, and this can enable the formation of a micro-colony. 
As a consequence, if the population of high bacteria is larger than the critical size $N_{th,4}$, then $\mathcal{G}^{(n)}$ always converges to a stable network with probability $1$.
\end{rem} 

\begin{figure}
     \centering
          \includegraphics[width=\figws]{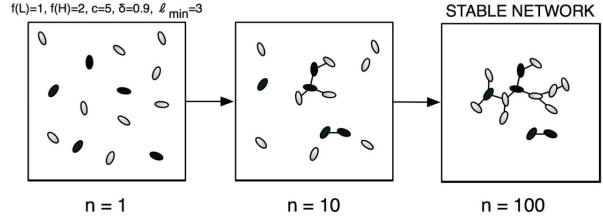}
\caption{Possible evolution of $\mathcal{G}^{(n)}$ in the incomplete information setting if $f(H) + \delta ( \ell_{min} - 1 ) f(H) + \delta (1 + \delta)f(L) \geq c(H)$. In this case $\mathcal{G}^{(n)}$ always converges to a stable network.}
\label{fig:para5}
\end{figure}

Fig. \ref{fig:para5} shows a possible evolution of a network $\mathcal{G}^{(n)}$ in the incomplete information setting with $\ell_{min} = 3$.
Even though a high type bacterium $i$ does not benefit from connecting with two low type bacteria, there is the possibility that other high type bacteria connect to $i$'s component before $i$ leaves it, and this enables the formation of a micro-colony. 
Notice that the system parameters in Fig. \ref{fig:para5} are the same as in Fig. \ref{fig:para3}, but in the complete information setting $\mathcal{G}^{(n)}$ does not converge to a stable network.

Finally, we investigate the robustness of a stable network.

\begin{theorem}
Let $\mathcal{G}^{(n)}$ a stable network, and assume that $L$ links are removed from it.

Consider the case $\ell_{min} = 1$.
If $f(H) + \delta ( \ell_{min} - 1 ) f(H) + \delta (1 + \delta)f(L) \geq c(H)$ then $\mathcal{G}^{(n)}$ converges again to a stable network, independently from the number $L$ of removed links.
Otherwise $\mathcal{G}^{(n)}$ converges again to a stable network if $L \leq \frac{K(H)}{2 N_{th,4}}$.

Consider the case $\ell_{min} \geq 2$.
If $K(H) \geq N_{th,4}-2$ and $K \geq N_{th,4}$, then $\mathcal{G}^{(n)}$ converges again to a stable network
\label{teo:inc_rob}
\end{theorem}

\begin{IEEEproof}
The case $\ell_{min} = 1$ is proven as in Theorem \ref{teo:com_rob}. 
The case $\ell_{min} \geq 2$ is proven using the same arguments as in the proof of Theorem \ref{teo:inc_conv}.
\end{IEEEproof}

\begin{rem}
If $\ell_{min} \geq 2$ and the high type bacteria number $K(H)$ is larger than the critical size $N_{th,4}$, then $\mathcal{G}^{(n)}$ converges again to a stable network, regardless from the number of broken links.
\end{rem}

\section{Simulations} 
\label{sec:sim} 

In this Section we present several illustrative results aimed to understand the essential characteristics of micro-colonies formation and their dependence on the key parameters of the system. 

We consider the basic settings represented in Table \ref{tab:1} and we run several series of simulations. 
In each series of simulations we vary the value of a single parameter, and for each value of the parameter we run $N_{sym} = 1000$ simulations to average the results. 
A single simulation consists of a maximum of $N_{slots} = 10^4$ time slots, if the network does not converge to a stable network before achieving the maximum number of time slots then the network is considered unstable.

\begin{table}
\begin{center}
\caption{Basic Simulation Settings.}
\begin{tabular}{|c|c|c|c|c|c|c|c|c|}
\textbf{Parameter description} & \textbf{Symbol} & \textbf{Value}\\
\hline
Number of bacteria & $K$ & $100$ \\
\hline
Ratio of high type bacteria & $\rho_H$ & $0.7$ \\
\hline
Spread factor & $\delta$ & 0.8 \\
\hline
Link cost for hight type bacteria & $c_H$ & 10 \\
\hline
Benefit received from high type bacteria & $f(H)$ & 2 \\
\hline
Benefit received from low type bacteria & $f(L)$ & 1 \\
\hline
Minimum link length & $\ell_{min}$ & 10 \\
\hline
Signaling parameter & m & $1$ \\ 
\hline
Maximum number of iterations & $N_{slots}$ & $10^4$ \\
\hline
Number of simulations for series & $N_{sym}$ & $1000$ \\
\hline
\end{tabular}
\label{tab:1}
\end{center}
\end{table}

\begin{figure}
\centering
\begin{subfigure}{.5\textwidth}
  \centering
  \includegraphics[width=0.95\linewidth]{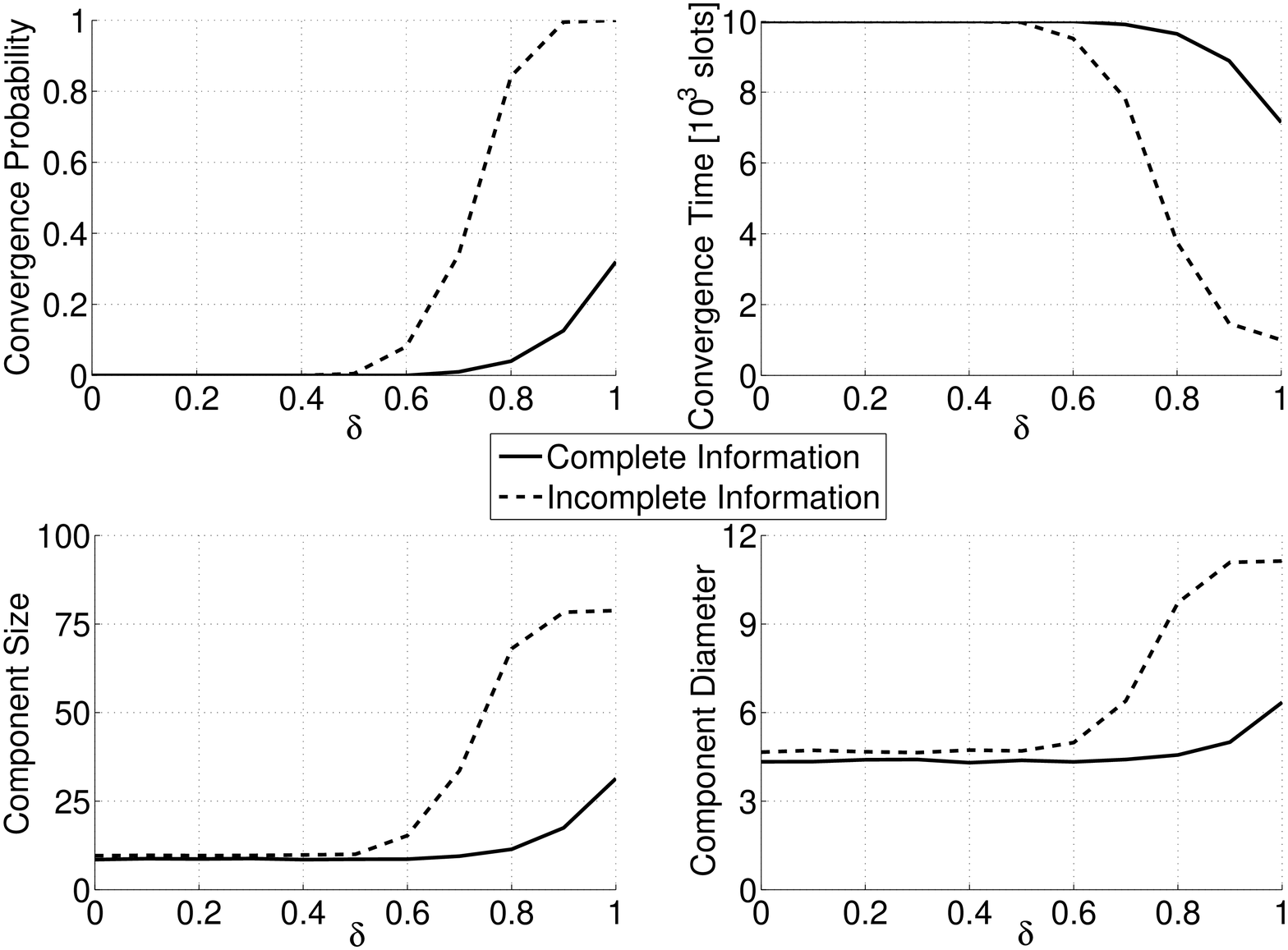}
  \caption{Spread factor.}
  \label{fig:simu1}
\end{subfigure}%
\begin{subfigure}{.5\textwidth}
  \centering
  \includegraphics[width=0.95\linewidth]{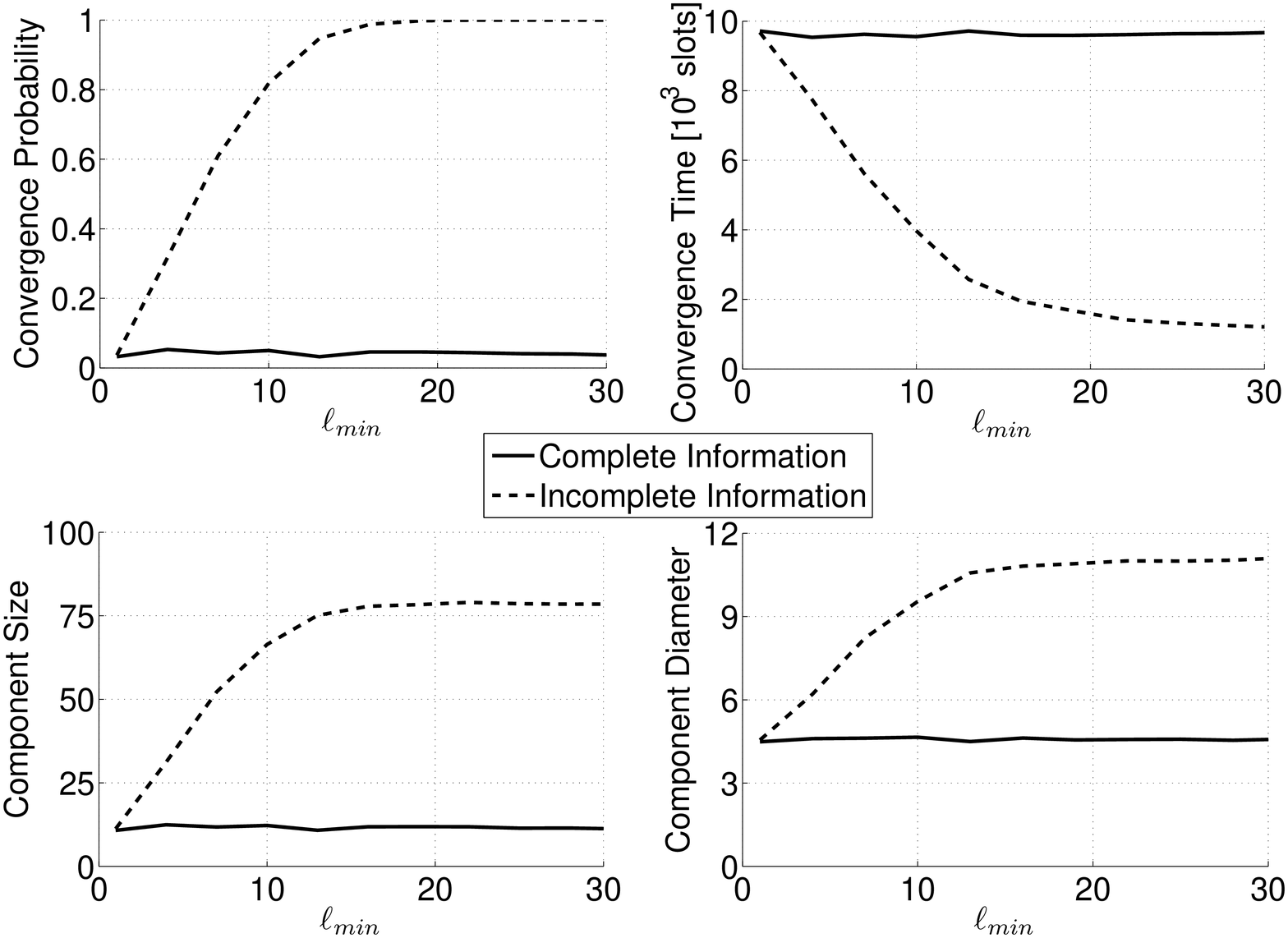}
  \caption{Minimum link length.}
  \label{fig:simu2}
\end{subfigure}
\caption{Convergence probability, convergence time, component size, and component diameter vs. (a) spread factor, and (b) minimum link length.}
\label{fig:test}
\end{figure}

We first analyze the impact of the spread factor $\delta$.
In Fig. \ref{fig:simu1} we plot the empirical convergence probability (top-left sub-figure), the average convergence time (top-right sub-figure), the average size of the largest component (bottom-left sub-figure), and the average diameter of the largest component (bottom-right sub-figure), for values of $\delta$ ranging from $0$ to $1$.
For both the complete and the incomplete information settings, if $\delta$ is very low then the network does not converge. 
Indeed, low type bacteria form micro-colonies anyway, but the largest micro-colony $\mathcal{C}$ has a size so small (about $10$) and it is so spread ($D_{\mathcal{C}}$ is about $5$) that in the complete information setting no high type bacteria has an incentive to join any micro-colonies, whereas in the incomplete information setting high type bacteria join temporarily some micro-colonies but then they eventually leave. 
The probability to converge to a stable network becomes positive for $\delta > 0.5$ in the incomplete information setting, and for $\delta > 0.8$ in the complete information setting. 
The convergence probability in the incomplete information setting is higher than the convergence probability in the complete information case because in the incomplete information case high type bacteria can join a micro-colony even if it is not currently beneficial for them, and this allow the micro-colony to grow faster and attract other high type bacteria.
Notice that, in this case, for $\delta \geq 0.9$ the network converges to a stable network with probability $1$.


Fig. \ref{fig:simu2} shows the impact of the minimum link length $\ell_{min}$.
The results for the complete information setting are not affected by $\ell_{min}$, indeed Lemma \ref{lem:com_act} proves that in this case bacteria always form stable links, regardless of $\ell_{min}$.
Unlike the complete information setting, the results in the incomplete information case are strongly influence by $\ell_{min}$.
On one hand, if $\ell_{min} = 1$ the incomplete information case coincides with complete information case, indeed bacteria forming non-beneficial links break the links immediately.
On the other hand, if $\ell_{min}$ is very large a high type bacterium waits for a long time before deciding whether to break its links, and this enables the formation of micro-colonies with high type bacteria.


\begin{figure}
\centering
\begin{subfigure}{.5\textwidth}
  \centering
  \includegraphics[width=0.95\linewidth]{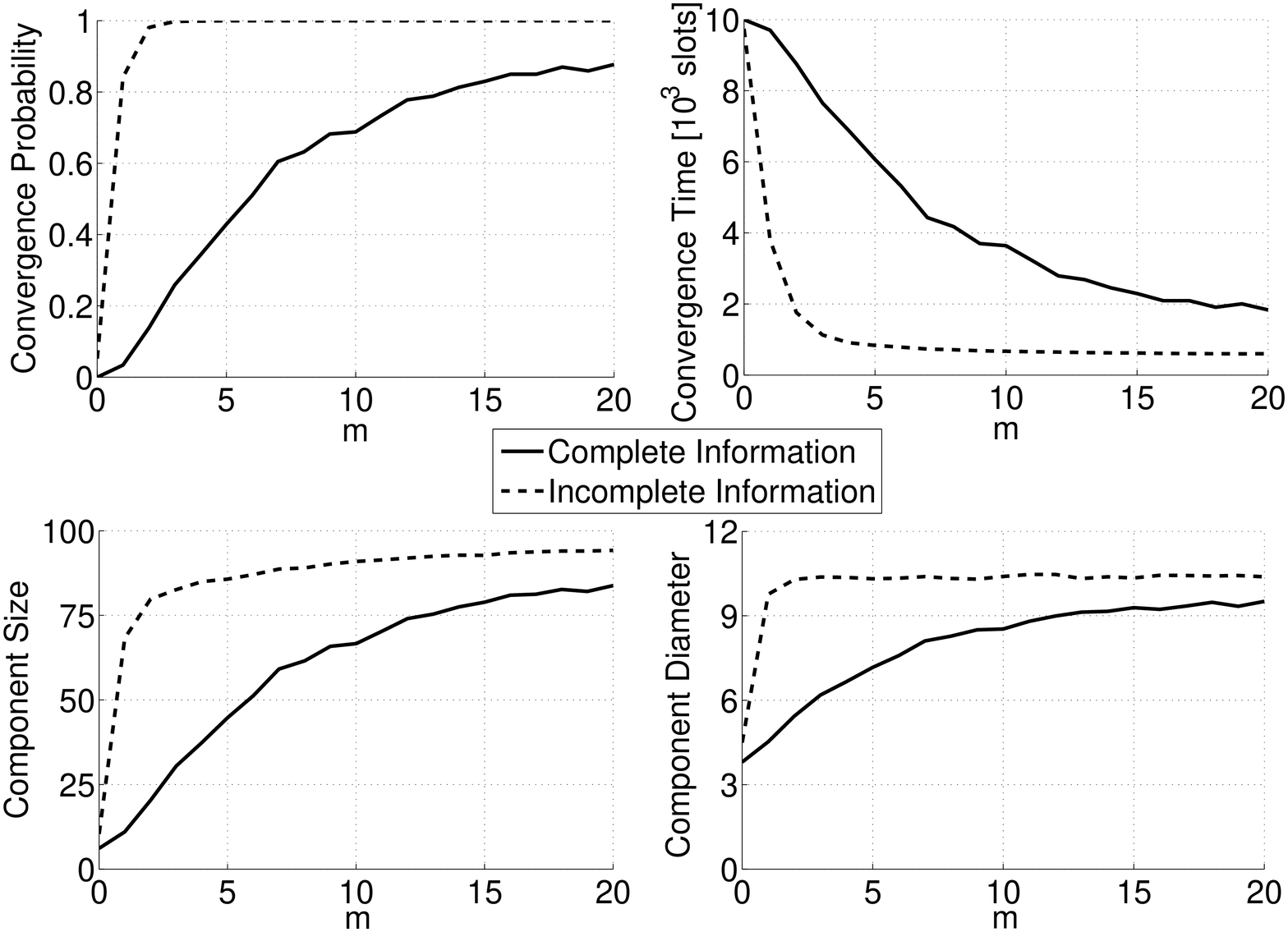}
  \caption{Signaling parameter.}
  \label{fig:simu3}
\end{subfigure}%
\begin{subfigure}{.5\textwidth}
  \centering
  \includegraphics[width=0.95\linewidth]{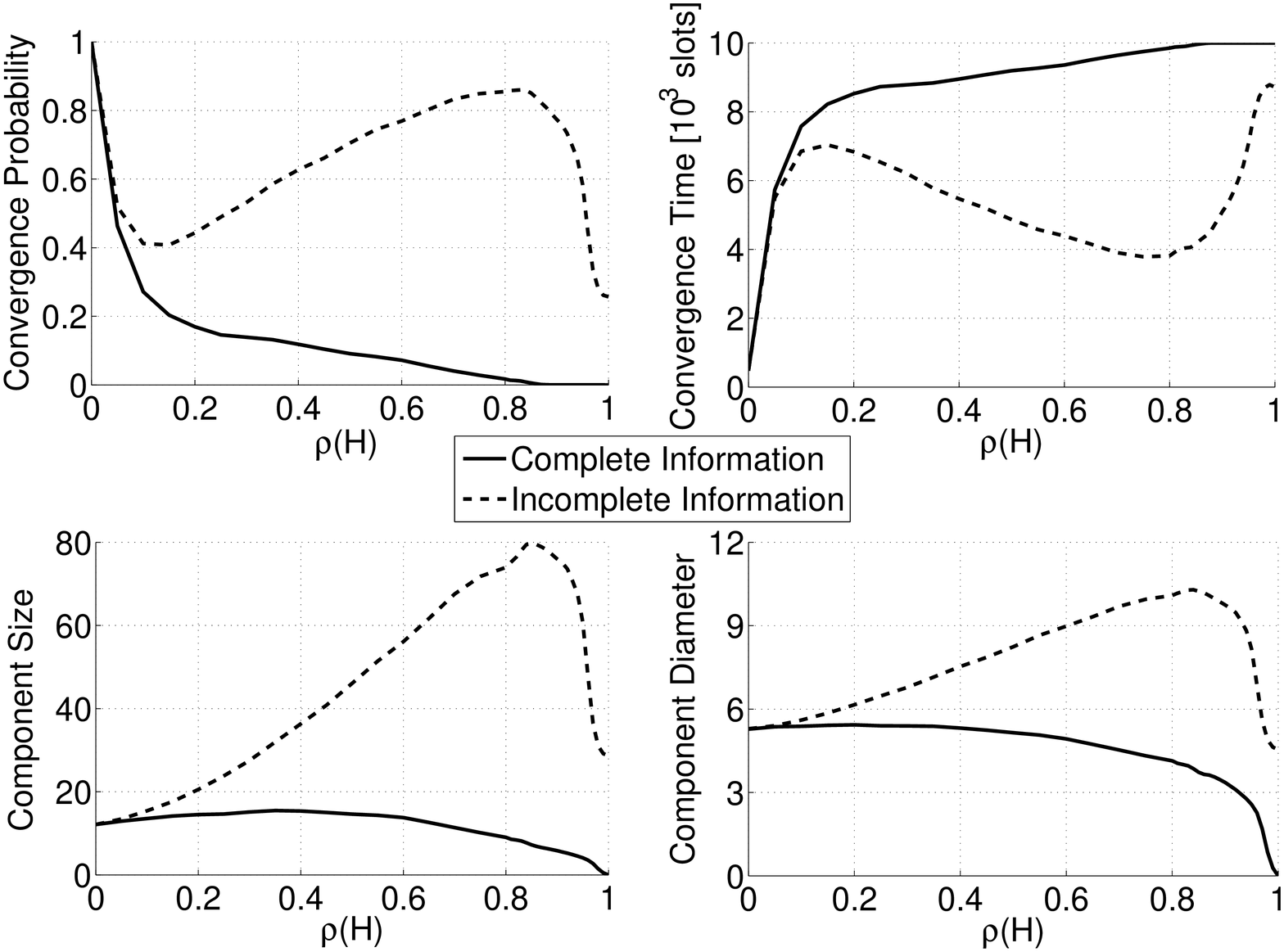}
  \caption{Ratio of high type bacteria.}
  \label{fig:simu4}
\end{subfigure}
\caption{Convergence probability, convergence time, component size, and component diameter vs. (a) signaling parameter, and (b) ratio of high type bacteria.}
\label{fig:test}
\end{figure}

Next we study the impact of the signaling mechanism.
We consider a linear signaling mechanism $h(x) = 1 + m x$, and we vary the \emph{signaling parameter} $m$ from $0$ to $20$: $m = 0$ means that the meetings are uniformly distributed, 
whereas the higher $m$ the higher the probability that a singleton bacterium is matched with a bacterium having many links.
Fig. \ref{fig:simu3} shows that the signaling mechanism has a positive effect in the formation of the micro-colonies: if $m$ is large, then the probability that the network converges is large, the convergence time is low, and the size of the component is large.  
Notice that the component diameter increases in $m$; however, this is due to the fact that the component size increases as well.
Since the component size increases with a much faster rate than the component diameter, then we can conclude that a large $m$ results in large and compact micro-colonies.


Fig. \ref{fig:simu4} evaluates the impact of the ratio of high type bacteria, $\rho_H$. 
If $\rho_H = 0$ the population is formed only by low type bacteria and the network converges to a stable network with probability $1$.
In the complete information, the higher $\rho_H$ the lower the probability that the network converges to a stable network, and as a consequence the lower the size of the largest component.
In the incomplete information the impact of $\rho_H$ is more complex. 
In fact, in the incomplete information settings two different factors are fundamentally important: the presence of low type bacteria that form the first micro-colonies, and the presence of high type bacteria that link to these micro-colonies and attract further bacteria.
As a consequence of this trade-off, there exists an optimal ratio of high type bacteria ($\rho_H = 0.85$) that maximizes the size of the largest component. 


\begin{figure}
     \centering
          \includegraphics[width=\figwss]{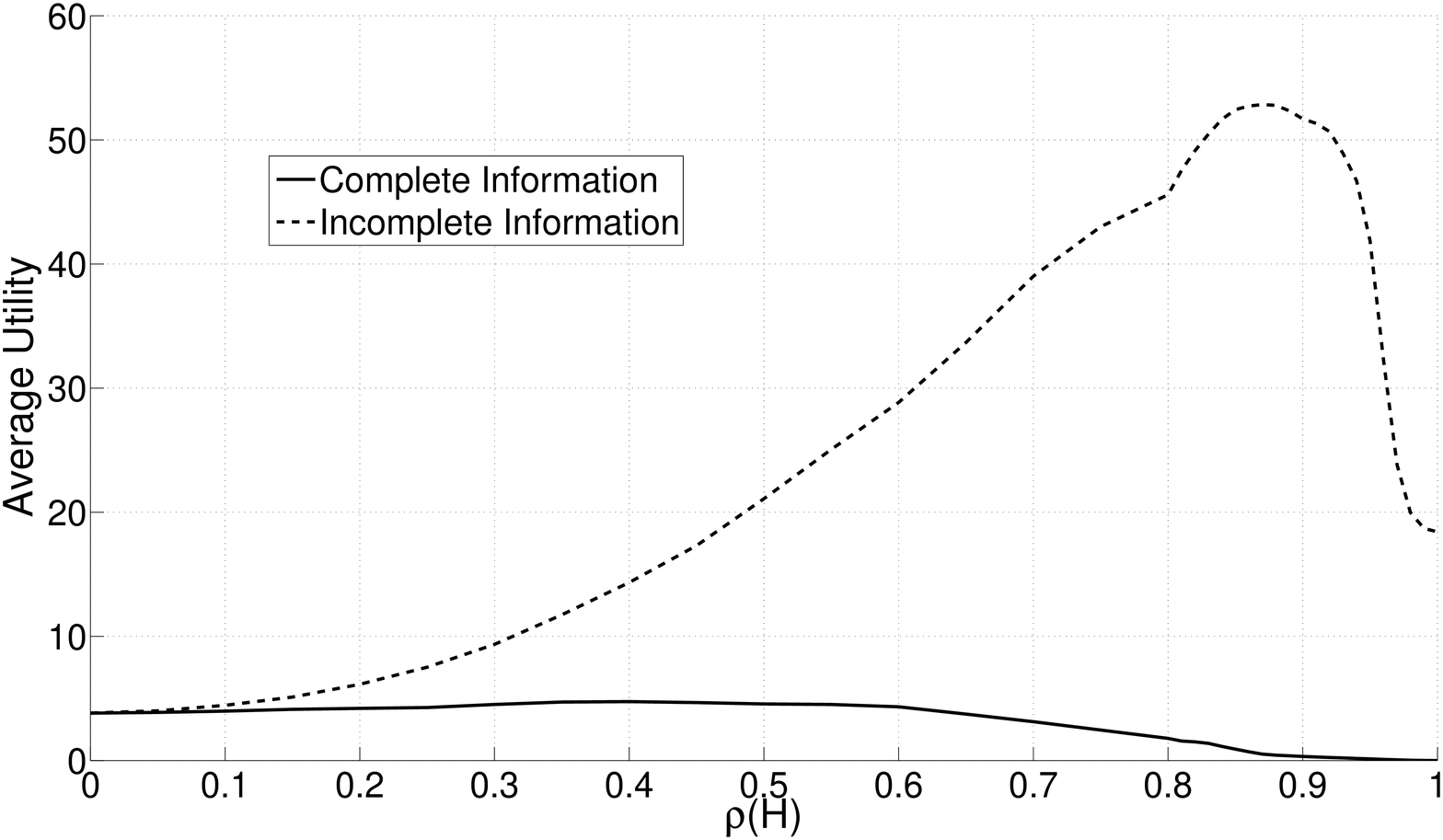}
\caption{Average utility for bacterium vs. ratio of high type bacteria.}
\label{fig:simu5}
\end{figure}

The trade-off between low and high type bacteria is even clearer in Fig. \ref{fig:simu5}, that represents the average utility that a bacterium obtains at the end of the simulation.
On one hand, a population of only low type bacteria always converges to a stable network, but the sizes of the micro-colonies are small and no bacterium adopts a higher production rate which would result in a larger benefit for all the bacteria belonging to the same micro-colony.
On the other hand, a population of only high type bacteria never converges to stable network, and the average utility is very low. 
There exists an optimal ratio of high type bacteria ($\rho_H = 0.87$) such that the average utility for bacterium is maximized. 
This implies that the long term benefit of a population of bacteria is maximized when the population is heterogeneous: both low and high type bacteria play a fundamental role in the formation of communities that exhibit enhanced antibiotic tolerance.


\section{Conclusion and Future Work} 
\label{sec:con} 

We proposed a parametrizable dynamic network formation game model to capture the dynamic interaction among bacteria in the formation of micro-colonies. 
We rigorously characterized some of the key properties of the network evolution depending on the parameters of the system, in both the complete and the incomplete information settings. 
In particular, we derived the conditions on the system parameters under which the network $\mathcal{G}^{(n)} $ converges to a stable network with probability $1$, under which $\mathcal{G}^{(n)}$ does not converge to a stable network with probability $1$, and under which the convergence of $\mathcal{G}^{(n)}$ is determined by chance.
Importantly, our study does not only characterize the properties of networks emerging in equilibrium, but it also provides important insights on how the network dynamically evolves and on how the formation history impacts the emerging networks in equilibrium. 
This analysis can be used to develop methods to influence on-the-fly the evolution of the network, and such methods can be useful to design biofilm therapeutic strategies.

As a continuation of this study, we plan to validate our model collecting experimental data. 
Using an approach similar to the one adopted to generate Fig. \ref{fig:para_intro}, we will record the whole history of formation of micro-colonies since the beginning of each experiment. 
We will use single Pseudomonas aeruginosa strains, as well as a mixture of different strains. 
The experiments will be performed in controlled environments to improve their repeatability, and will be repeated for several times to obtain a statistically significant number of data. 
The collected data will be used to tune the parameters of our model and to compare the real results with the results predicted by our model.

\appendices

\bibliographystyle{IEEEtran}
\bibliography{IEEEabrv,biblio}

\begin{thebibliography}{10}
\providecommand{\url}[1]{#1}
\csname url@samestyle\endcsname
\providecommand{\newblock}{\relax}
\providecommand{\bibinfo}[2]{#2}
\providecommand{\BIBentrySTDinterwordspacing}{\spaceskip=0pt\relax}
\providecommand{\BIBentryALTinterwordstretchfactor}{4}
\providecommand{\BIBentryALTinterwordspacing}{\spaceskip=\fontdimen2\font plus
\BIBentryALTinterwordstretchfactor\fontdimen3\font minus
  \fontdimen4\font\relax}
\providecommand{\BIBforeignlanguage}[2]{{%
\expandafter\ifx\csname l@#1\endcsname\relax
\typeout{** WARNING: IEEEtran.bst: No hyphenation pattern has been}%
\typeout{** loaded for the language `#1'. Using the pattern for}%
\typeout{** the default language instead.}%
\else
\language=\csname l@#1\endcsname
\fi
#2}}
\providecommand{\BIBdecl}{\relax}
\BIBdecl

\bibitem{Stoodley2002}
P.~Stoodley, K.~Sauer, D.~G. Davies, and J.~W. Costerton, ``Biofilms as complex
  differentiated communities,'' \emph{Annual Review of Microbiology}, vol.~56,
  pp. 187--209, 2002.

\bibitem{Parsek03}
M.~R. Parsek and P.~K. Singh, ``Bacterial biofilms: An emerging link to disease
  pathogenesis,'' \emph{Annual Review of Microbiology}, vol.~57, pp. 677--701,
  2003.

\bibitem{Costerton99}
J.~W. Costerton, P.~S. Stewart, and E.~P. Greenberg, ``Bacterial biofilms: a
  common cause of persistent infections,'' \emph{Science}, vol. 284, pp.
  1318--1322, 1999.

\bibitem{Drenkard03}
E.~Drenkard, ``Antimicrobial resistance of {P}seudomonas aeruginosa biofilms,''
  \emph{Microbes Infect}, vol.~5, pp. 1213--1219, 2003.

\bibitem{Gerard11}
K.~M. Colvin, V.~D. Gordon, K.~Murakami, B.~R. Borlee, D.~J. Wozniak, G.~C.~L.
  Wong, and M.~R. Parsek, ``The pel polysaccharide can serve a structural and
  protective role in the biofilm matrix of {P}seudomonas aeruginosa,''
  \emph{PLoS Pathog}, vol.~7, no.~1, p. e1001264, 01 2011.

\bibitem{Gerard13}
K.~Zhao, B.~S. Tseng, B.~Beckerman, F.~Jin, M.~L. Gibiansky, J.~J. Harrison,
  E.~Luijten, M.~R. Parsek, and G.~C.~L. Wong, ``Psl trails guide exploration
  and microcolony formation in {P}seudomonas aeruginosa biofilms,''
  \emph{Nature}, vol. 497, no. 7449, pp. 388--391, May 2013.

\bibitem{Watts01}
A.~Watts, ``A dynamic model of network formation,'' \emph{Games and Economic
  Behavior}, vol.~34, pp. 331--341, 2001.

\bibitem{Velicer03}
G.~J. Velicer, ``Social strife in the microbial world,'' \emph{Trends
  Microbiol}, vol.~11, pp. 330--337, 2003.

\bibitem{Shrout06}
J.~D. Shrout, D.~L. Chopp, C.~L. Just, M.~Hentzer, M.~Givskov, and M.~R.
  Parsek, ``The impact of quorum sensing and swarming motility on {P}seudomonas
  aeruginosa biofilm formation is nutritionally conditional,'' \emph{Mol.
  Microbiol.}, vol.~62, pp. 1264--1277, 2006.

\bibitem{Moore09}
M.~Moore, T.~Suda, and K.~Oiwa, ``Molecular communication: Modeling noise
  effects on information rate,'' \emph{{IEEE} Trans. Nanobiosci.}, vol.~8,
  no.~2, pp. 169--180, 2009.

\bibitem{Pierobon11}
M.~Pierobon and I.~Akyildiz, ``Noise analysis in ligand--binding reception for
  molecular communication in nanonetworks,'' \emph{{IEEE} Trans. Signal
  Process.}, vol.~59, no.~9, pp. 4168--4182, 2011.

\bibitem{Pierobon13}
M.~Pierobon and I.~F. Akyildiz, ``Capacity of a diffusion--based molecular
  communication system with channel memory and molecular noise,'' \emph{{IEEE}
  Trans. Inf. Theory}, vol.~59, no.~2, pp. 942--954, 2013.

\bibitem{Eckford11}
N.~Farsad, A.~Eckford, S.~Hiyama, and Y.~Moritani, ``A simple mathematical
  model for information rate of active transport molecular communication,'' in
  \emph{Proc. IEEE INFOCOM Workshop Mol. Nanoscale Commun.}, 2011, pp.
  473--478.

\bibitem{Eckford12_1}
K.~V. Srinivas, A.~Eckford, and R.~Adve, ``Molecular communication in fluid
  media: The additive inverse gaussian noise channel,'' \emph{{IEEE} Trans.
  Inf. Theory}, vol.~58, no.~7, pp. 4678--4692, 2012.

\bibitem{Atakan12}
B.~Atakan and O.~B. Akan, ``Nanoscale communication with molecular arrays in
  nanonetworks,'' \emph{{IEEE} Trans. Nanobiosci.}, pp. 149--160, 2012.

\bibitem{Jackson96}
M.~O. Jackson and A.~Wolinsky, ``A strategic model of social and economic
  networks,'' \emph{Journal of Economic Theory}, vol.~71, pp. 44--74, 1996.

\bibitem{Goyal00}
V.~Bala and S.~Goyal, ``A noncoorperative model of network formation,''
  \emph{Econometrica}, vol.~68, no.~5, pp. 1181--1229, 2000.

\bibitem{Johnson00}
C.~Johnson and R.~P. Gilles, ``Spatial social networks,'' \emph{Review of
  Economic Design}, vol.~5, no.~3, pp. 273--299, 2000.

\bibitem{Haller03}
H.~Haller and S.~Sarangi, ``Nash networks with heterogeneous agents,''
  \emph{Discussion Papers of DIW Berlin}, 2003.

\bibitem{Galeotti06}
A.~Galeotti, ``One-way flow networks: The role of heterogeneity,''
  \emph{Economic Theory}, vol.~29, no.~1, pp. 163--179, 2006.

\bibitem{Goyal06}
A.~Galeotti, S.~Goyal, and J.~Kamphorst, ``Network formation with heterogeneous
  players,'' \emph{Games and Economic Behavior}, vol.~54, pp. 353--372, 2006.

\bibitem{Goyal10}
A.~Galeotti and S.~Goyal, ``The law of the few,'' \emph{The American Economic
  Review}, vol. 100, no.~4, pp. 1468--1492, 2010.

\bibitem{Zhang12}
Y.~Zhang and M.~van~der Schaar, ``Information production and link formation in
  social computing systems,'' \emph{{IEEE} J. Sel. Areas Commun.}, vol.~30,
  no.~11, pp. 2136--2145, 2012.

\bibitem{Koenig12}
M.~D. Koenig, C.~J. Tessone, and Y.~Zenou, ``Nestedness in networks: A
  theoretical model and some applications,'' \emph{Stanford Institute for
  Economic Policy Research Discussion Papers}, 2012.

\bibitem{Zhang13}
Y.~Zhang and M.~van~der Schaar, ``Information dissemination and link formation
  among self-interested agents,'' \emph{{IEEE} J. Sel. Areas Commun.}, vol.~31,
  no.~6, pp. 1115--1123, 2013.

\bibitem{WestGraphBook}
D.~B. West, \emph{Introduction to Graph Theory (2nd Edition)}.\hskip 1em plus
  0.5em minus 0.4em\relax Prentice Hall, Aug. 2000.

\bibitem{Guzman95}
L.~M. Guzman, D.~Belin, M.~J. Carson, and J.~Beckwith, ``Tight regulation,
  modulation, and high-level expression by vectors containing the arabinose
  {PBAD} promoter,'' \emph{J Bacteriol}, vol. 177, pp. 4121--4130, 1995.

\bibitem{Mah03}
T.~F. Mah, B.~Pitts, B.~Pellock, G.~C. Walker, P.~S. Stewart, and G.~A.
  {O'Toole}, ``A genetic basis for {P}seudomonas aeruginosa biofilm antibiotic
  resistance,'' \emph{Nature}, vol. 426, no. 6964, pp. 306--310, 2003.

\bibitem{Gibbs08}
K.~A. Gibbs, M.~L. Urbanowski, and E.~P. Greenberg, ``Genetic determinants of
  self identity and social recognition in bacteria,'' \emph{Science}, pp.
  256--259, Jul. 2008.

\bibitem{Toole00}
G.~A. {O'Toole}, H.~B. Kaplan, and R.~Kolter, ``Biofilm formation as microbial
  development,'' \emph{Annual Review of Microbiology}, vol.~54, no.~1, pp.
  49--79, 2000.

\end{thebibliography}

\end{document}